\documentclass[review, 3p, sort&compress]{elsarticle}

\usepackage{amsmath, amsfonts, amssymb}
\usepackage{accents}
\usepackage{color}
\usepackage{mathrsfs}
\usepackage{graphicx}
\usepackage{multirow}
\usepackage{subfigure}
\usepackage{url}
\usepackage{lineno}

\usepackage[normalem]{ulem}

\renewcommand{\d}{\mathrm{d}}

\newcommand{\td}[2]{\frac{\d #1}{\d #2}}

\newcommand{\pd}[2]{\frac{\partial #1}{\partial #2}}
\newcommand{\pdf}[2]{\partial #1/\partial #2}
\newcommand{\pdd}[2]{\frac{\partial^2 #1}{\partial #2^2}}

\newcommand{\E}[1]{\times 10^{#1}}

\newcommand{\etal}{\emph{et al.~}}
\newcommand{\unit}[1]{$^{#1}$}

\definecolor{darkgreen}{rgb}{0.0, 0.26, 0.15}

\newenvironment{subeq}[1]{
\begin{subequations}
#1
\end{subequations}
}

\begin{document}

\begin{frontmatter}

\title{Modelling ultra-fast nanoparticle melting with the Maxwell--Cattaneo equation}

\author[CRM,OX]{Matthew~G.~Hennessy\corref{cor1}}
\ead{hennessy@maths.ox.ac.uk}

\author[CRM,UPC]{Marc~Calvo-Schwarzw\"alder}
\ead{mcalvo@crm.cat}

\author[CRM,UPC]{Timothy~G.~Myers}
\ead{tmyers@crm.cat}

\cortext[cor1]{Corresponding author}

\address[CRM]{Centre de Recerca Matem\`{a}tica, Campus de Bellaterra, Edifici C, 08193 Bellaterra (Barcelona), Spain}
\address[OX]{Mathematical Institute, University of Oxford, Andrew Wiles Building, Woodstock Road, Oxford OX2 6GG, United Kingdom}
\address[UPC]{Departament de Matem\`{a}tiques, Universitat Polit\`{e}cnica de Catalunya, 08028 Barcelona, Spain}

\begin{abstract}
  The role of thermal relaxation in nanoparticle melting is studied using a mathematical model based on the Maxwell--Cattaneo equation for heat conduction. The model is formulated in terms of a two-phase Stefan problem. We consider the cases of the temperature profile being continuous or having a jump across the solid-liquid interface. The jump conditions are derived from the sharp-interface limit of a phase-field model that accounts for variations in the thermal properties between the solid and liquid. The Stefan problem is solved using asymptotic and numerical methods. The analysis reveals that the Fourier-based solution can be recovered from the classical limit of zero relaxation time when either boundary condition is used.  However, only the jump condition avoids the onset of unphysical ``supersonic'' melting, where the speed of the melt front exceeds the finite speed of heat propagation. These results conclusively demonstrate that the jump condition, not the continuity condition, is the most suitable for use in models of phase change based on the Maxwell--Cattaneo equation. Numerical investigations show that thermal relaxation can increase the time required to melt a nanoparticle by more than a factor of ten. Thus, thermal relaxation is an important process to include in models of nanoparticle melting and is expected to be relevant in other rapid phase-change processes.
\end{abstract}


\end{frontmatter}

%%%%%%%%%%%%%%%%%%%%%%%%%%%%%%%%%%%%%%%%%%%%%%%%%%%%%%%
\section{Introduction}
%Nanotechnology has become a very important area of research due to the numerous applications it has to offer \cite{salata2004,mah2000,nam2003,garnett2011,pop2006,ge2012}. In biology and medicine \cite{salata2004}, for instance, it is used in drug and gene delivery \cite{mah2000}, protein detection \cite{nam2003}, and tissue engineering \cite{ma2003}. Industrial applications, such as the nanowire-based solar cells \cite{garnett2011}, nanoelectric devices \cite{pop2006} or improving the life cycle of ion batteries by means of nanowires \cite{ge2012} often involve high temperatures and a wrong description of the heat transfer mechanisms may lead to melting or device failure \cite{nie2011}. It is therefore crucial to be able to correctly model heat transfer and phase change processes at small scales to improve the applicability of nanotechnology in the future.

% The classical mathematical model describing a phase change process is based on the Fourier's heat conduction law and is known as the Stefan problem; see \cite{gupta2003,alexiades1992} \mh{Sentence not relevant for this paper}. Although this formulation has been successfully used to understand these mechanisms \mh{what mechanisms?} at large scales, it cannot be applied at the nanoscale. 

Technological advancements are driving the need to better understand the thermal characteristics of nanosystems \cite{cahill2003, cahill2014}.  Perhaps the simplest and most theoretically studied nanosystem is the spherical nanoparticle.  The one-dimensional geometry of a nanoparticle is ideal for developing mathematical models which can be compared against experimental data and used to assess new theories of nanoscale heat transport and phase change.  From a practical perspective, nanoparticles play fundamental roles in novel drug delivery systems \cite{yih2006}, materials with modified properties \cite{dieringer2006,grohn2001}, and for improving the efficiency of solar collectors \cite{cregan2015}. Many current and future applications of nanoparticles require quantitative knowledge of how they respond to their thermal environment and their behaviour during melting.

%Nanoparticles play an important role in nanotechnology due to the large number of applications that it has to offer. These range from biology and medicine \cite{ahmad2012,salata2004}, where they are used for diagnosis \cite{mornet2004}, drug delivery \cite{yih2006} or cancer therapy \cite{jain2012}, to industry \cite{stark2015}, where nanoparticles are used in the manufacturing of optoelectronic devices \cite{tanabe2007} and novel materials with modified properties \cite{dieringer2006,grohn2001} or to improve the efficiency of solar collectors \cite{cregan2015}, for example. Some of these applications often involve large amounts of energy and high temperatures that can potentially lead to melting or device failure \cite{nie2011} and therefore a detailed understanding of how nanoparticles respond to changes in their thermal environment is of vital importance.

The thermal response of a nanoparticle differs from that of a macroscopic body for two main reasons. Firstly, the large ratio of surface to bulk atoms leads to many key thermophysical properties, such as melt temperature \cite{buffat1976,david1995,wronski1967}, latent heat \cite{lai1996, zhang2000, sun2007}, and surface energy \cite{tolman1949} becoming dependent on the size of the nanoparticle. Secondly, the mechanism of thermal transport changes between the macroscale and the nanoscale. At the macroscale, heat is transported by a diffusive process that is driven by frequent collisions between thermal energy carriers known as phonons. Diffusive transport of heat across macroscopic length scales is well described by Fourier's law. On nanometer length scales, thermal energy is transported by a ballistic process driven by infrequent collisions between phonons. The finite time between phonon collisions results in a wave-like propagation of heat with finite speed \cite{joseph1989, joseph1990}. Since Fourier's law leads to an infinite speed of heat propagation, it is not suitable for describing ballistic energy transport.

Recent theoretical studies of nanoparticle melting have investigated the role of size-dependent material properties \cite{back2014, font2013, font2015, mccue2009, myers2015, ribera2016, myers2016} using models that are derived from Fourier's law. Consequently, these models can only capture the diffusive regime of thermal transport rather than the ballistic regime that is more relevant for nanosystems. These studies also predict that nanoparticle melting occurs on pico- to nanosecond time scales, which are of the same order as the time between phonon collisions in many metallic systems \cite{joseph1989}. Thus, there is strong evidence to suggest that non-Fourier heat conduction should be accounted for in theoretical models of nanoparticle melting.

The observation of wave-like thermal transport in liquid helium \cite{peshkov1946} prompted the derivation of new models of heat conduction that avoid the infinite speed of heat propagation predicted by Fourier's law. Cattaneo \cite{cattaneo1958} and Vernotte \cite{vernotte1958} are often credited with proposing the first non-Fourier model of heat conduction by accounting for the thermal relaxation time of a material. However, similar equations were, in fact, proposed decades earlier \cite{sobolev2018}. Microscopically, the relaxation time reflects the time between phonon collisions; macroscopically, it represents a time lag between the imposition of a temperature gradient and the creation of a thermal flux. The model proposed by Cattaneo and Vernotte is sometimes called the Maxwell--Cattaneo (or Maxwell--Cattaneo--Vernotte) equation due to the similarity with the equation derived by the British physicist when providing a mathematical basis for the kinetic theory of gases \cite{maxwell1867}. When the Maxwell--Cattaneo equation is combined with conservation of energy, the so-called hyperbolic heat equation for the temperature is obtained. The Maxwell--Cattaneo equation predicts that thermal energy is transported as a wave with finite speed, making it better suited than Fourier's law for capturing the ballistic transport regime \cite{chester1963}. However, it can also lead to results which violate the laws of thermodynamics \cite{bright2009}. Extensions to the classical theory of irreversible thermodynamics have been proposed as a means of rationalising this issue \cite{jou2010} and used to derive a family of heat conduction models that generalise the Maxwell--Cattaneo equation \cite{van2012}.

The first studies that incorporated the Maxwell--Cattaneo equation into models of phase change focused on mathematical issues \cite{colli1993, friedman1989, showalter1987} such as the correct form of boundary condition at the interface where phase change occurs \cite{glass1991, greenberg1987}. Some authors proposed that the temperature should be continuous across the interface \cite{friedman1989, solomon1985}, while others suggested that there should be jump in temperature due to the hyperbolic nature of the governing equations  \cite{greenberg1987, sobolev1991, sobolev1996}. Numerical studies based on one-dimensional Cartesian geometries showed that non-Fourier heat conduction affects the dynamics of phase change on time scales that are commensurate with the relaxation time of the material \cite{sadd1977, liu2009}. Subsequent studies used the Maxwell--Cattaneo equation to model rapid solidification problems \cite{sobolev2015ii}, dendrite formation in metal baths \cite{mullis1997}, pulsed-laser surface treatment \cite{wang2000}, cryopreservation \cite{deng2003, ahmadikia2012}, and cryosurgery of lung cancer \cite{kumar2017}. In many of these works, the governing equations are solved using numerical methods. In fact, it appears that little attention has been given to seeking analytical or asymptotic solutions to non-Fourier models of phase change. Sobolev \cite{sobolev1991, sobolev1996} couples non-Fourier models of heat transfer to the Stefan problem and derives analytical solutions for the case of constant interface velocities. Hennessy \etal\cite{hennessy2018} carried out a detailed asymptotic analysis of a non-Fourier model of solidification that accounts for the mean free path of phonons.  The authors showed that non-classical transport mechanisms lead to additional time regimes, each with distinct solidification kinetics.

% Recently, some authors have used extensions of the Maxwell--Cattaneo equation to explicitly account for spatial interactions between heat carriers in their formulation of nanoscale phase change \cite{sobolev2014,hennessy2018}. However, in the context of melting nanoparticles, where phase change is predominantly driven by heat conduction through the liquid, it is non-trivial to account for spatial interactions between heat carriers as the theoretical framework for liquids is still being developed \cite{bolmatov2012}.

In this paper, we investigate the role of non-Fourier heat conduction and thermal relaxation in nanoparticle melting using a mathematical model based on the Maxwell--Cattaneo equation. Furthermore, we address the issue of which boundary condition should be imposed at the solid-liquid interface. A new form of the temperature jump condition is systematically derived from the sharp-interface limit of a non-Fourier phase-field model that accounts for variations in the material properties of the solid and liquid. We use a combination of asymptotic and numerical methods to study the dynamics of phase change. The analysis uncovers a key shortcoming of the continuity boundary condition for the temperature, which results in the speed of the melt front exceeding the finite speed of heat propagation in the materiasls. An asymptotically reduced model is obtained in the limit of large Stefan number. To the best of our knowledge, this is the first time that such a reduction has been carried out on a model that is based on the Maxwell--Cattaneo equation. Finally, we assess the validity of our model in the context of classical thermodynamics.

The paper is organised as follows. In Sec.~\ref{sec:model}, a theoretical model for spherical nanoparticle melting is developed. The numerical scheme is presented in Sec.~\ref{sec:numerics}. Asymptotic solutions under different limiting situations are computed in Sec.~\ref{sec:asymptotics}. A discussion based on the results then follows in Sec.~\ref{sec:discussion}. The paper concludes in Sec.~\ref{sec:conclusion}.

%-----------------------------------------------------
\section{Model}\label{sec:model}
\begin{figure}
	\centering
    \includegraphics[width=.3\textwidth]{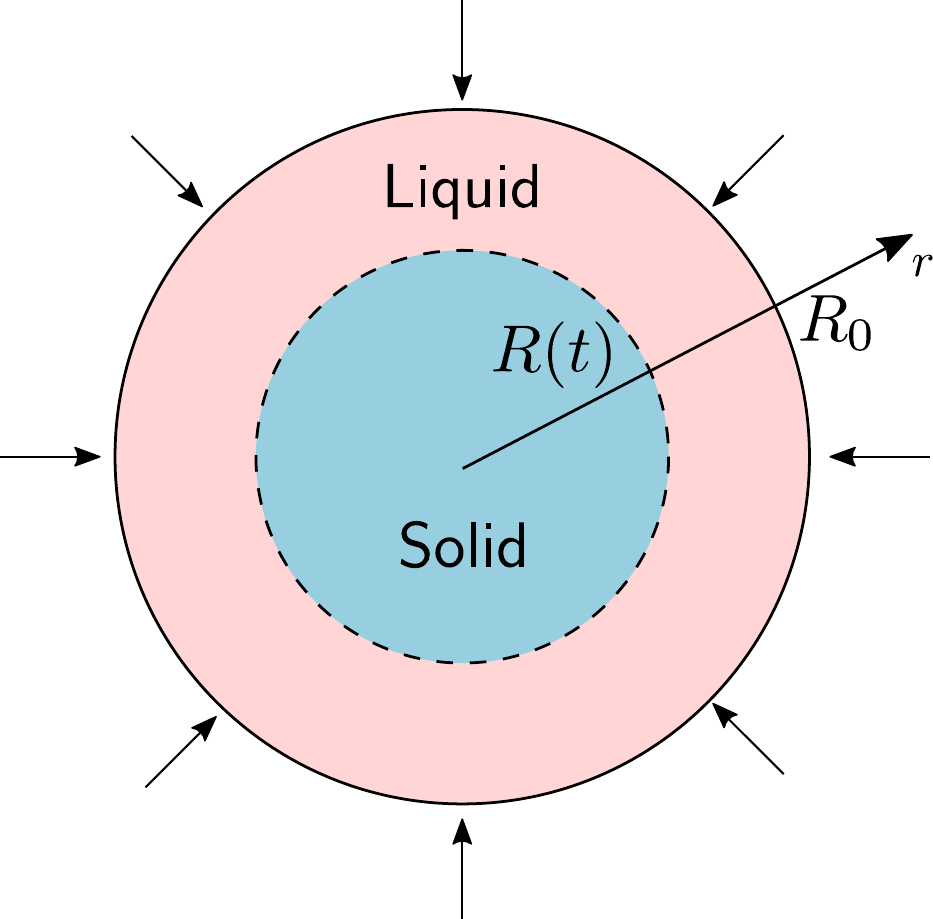}
    \caption{Spherically symmetry nanoparticle melting.  The melting process is driven by an influx of thermal energy from the environment. The combined radius of the solid-liquid system is $R_0$ and assumed to be constant.  The position of the solid-liquid interface is denoted by $R(t)$.}    
    \label{fig:schematic}
\end{figure}

We consider the transient melting of a spherical nanoparticle of radius $R_0$ that is suspended in a warm environment held at temperature $T_e$. This environment is envisioned to be a solid matrix that the nanoparticle is embedded in \cite{sheng1998} or an inert gaseous atmosphere \cite{sun2007}. Back \etal\cite{back2014} consider the case where the solid core of the nanoparticle is surrounded by an infinite liquid bath. The transfer of heat from the environment into the nanoparticle initiates and sustains melting.  The entire melting process is assumed to be symmetric so that it is sufficient to consider a one-dimensional model involving time $t$ and the spherical radial coordinate $r$. The radial position of the solid-liquid interface, or melt front, is given by $r = R(t)$. Figure \ref{fig:schematic} illustrates this situation.

To enable the identification of the dominant mechanisms of phase change, the density variation that occurs during melting is assumed to be so small that the change in total volume of the liquid-solid system and the motion of liquid can be neglected. Density-driven motion and expansion of the liquid are considered in Refs.~\cite{font2015, ribera2016}.  We also assume for simplicity that the specific heat of the liquid and solid are equal. As will be shown in Sec.~\ref{sec:parameters}, these assumptions are reasonable for many metals.

\subsection{Bulk equations}
% The temperature $T_s$ and thermal flux $q_s$ of the solid are assumed to be
% \begin{align}
%   T_s(r,t) = T_m(R), \quad q_s(r,t) = 0.
%   \label{eqn:solid}
% \end{align}
% \begin{align}
%   \rho_l \pd{u_l}{t} + \frac{1}{r^2}\pd{}{r}\left(r^2 q_l \right) = 0,
%   \label{eqn:energy}
% \end{align}
The governing equations are derived by considering the local form of energy conservation in the solid and liquid phases:
\subeq{
  \label{eqn:comp_energy}
  \begin{align}
    \rho \pd{u_s}{t} + \frac{1}{r^2}\pd{}{r}\left(r^2 q_s\right) &= 0, \qquad 0 < r < R(t), \\
    \rho \pd{u_l}{t} + \frac{1}{r^2}\pd{}{r}\left(r^2 q_l\right) &= 0, \qquad R(t) < r < R_0,
  \end{align}
}
where $\rho$ is the mean density of the solid and liquid phase, $u_s$ and $u_l$ are the internal energy per unit mass of the solid and liquid, and $q_s$ and $q_l$ are thermal fluxes. The conduction of heat is described by the Maxwell--Cattaneo equation. Thus, the thermal fluxes satisfy
\subeq{
  \label{eqn:flux}
  \begin{align}
    \tau_s \pd{q_s}{t} + q_s &= -k_s\pd{T_s}{r}, \qquad 0 < r < R(t), \\
    \tau_r \pd{q_l}{t} + q_l &= -k_l \pd{T_l}{r}, \qquad R(t) < r < R_0,
  \end{align}
}
where $\tau_s$ and $\tau_s$ are thermal relaxation times of the solid and liquid and $k_s$ and $k_l$ are thermal conductivities. The first term on the left-hand sides of \eqref{eqn:flux} captures the delayed response of the flux to a change in thermal environment.  This delay is caused by the finite time between the interactions of heat carriers.  The introduction of a time derivative in \eqref{eqn:flux} leads to the flux depending on the history, \emph{i.e.} the time integral, of the temperature gradient. For times that are large compared to the relaxation time, the memory term is negligible and Fourier's law is recovered.

The internal energy of the solid and liquid per unit mass can be defined as
\begin{align}
\label{eqn:u}
 u_s = c (T_s - T_m^*), \quad
 u_l = c (T_l - T_m^*) + L_m,
\end{align}
where $c$ is the mean heat capacity of the solid and liquid at constant pressure, $T_s$ and $T_l$ are the temperatures of the solid and liquid, respectively; $T_m^*$ is the bulk melting temperature; and $L_m$ is the latent heat of melting. By substituting \eqref{eqn:u} into the local energy balances \eqref{eqn:comp_energy}, evolution equations for the temperature are obtained:
\subeq{\label{eqn:comp_T}
  \begin{align}
    \rho c \pd{T_s}{t} + \frac{1}{r^2}\pd{}{r}\left(r^2 q_s\right) &= 0, \qquad 0 < r < R(t), \\
    \rho c \pd{T_l}{t} + \frac{1}{r^2}\pd{}{r}\left(r^2 q_l\right) &= 0, \qquad R(t) < r < R_0,
  \end{align}
}
By differentiating each equation in \eqref{eqn:comp_T} with respect to time and using the Maxwell--Cattaneo equations \eqref{eqn:flux}, the temperatures can be shown to satisfy a pair of hyperbolic heat equations given by
\subeq{
  \label{eqn:hhe}
  \begin{align}
    \tau_s \pdd{T_s}{t} + \pd{T_s}{t} &= \frac{\kappa_s}{r^2}\pd{}{r}\left(r^2\pd{T_s}{r}\right), \qquad 0 < r < R(t), \\
    \tau_l \pdd{T_l}{t} + \pd{T_l}{t} &= \frac{\kappa_l}{r^2}\pd{}{r}\left(r^2\pd{T_l}{r}\right), \qquad R(t) < r < R_0, 
  \end{align}
}
where $\kappa_s = k_s / (\rho c)$ and $\kappa_l = k_l / (\rho c)$ are the thermal diffusivity of the solid and liquid, respectively.  Unlike the standard parabolic heat equation, the hyperbolic heat equation predicts that thermal energy is transported in a wave-like manner with a finite speed given by $(\kappa_i / \tau_i)^{1/2}$. For $t \gg \tau_i$, or $\tau_i \to 0$, the hyperbolic heat equations reduce to the standard heat equation. Despite the elegant form of the hyperbolic heat equations given by \eqref{eqn:hhe}, it is advantageous to work with the first-order hyperbolic system given by \eqref{eqn:flux} and \eqref{eqn:comp_T} because the boundary conditions will depend on the fluxes $q_s$ and $q_l$.

To facilitate the derivation of the boundary conditions, the solid-liquid interface $r = R(t)$ can be considered as a shock where the solutions to the governing equations exhibit a discontinuity. Under this interpretation, it is possible to define an internal energy $u$, temperature $T$, and flux $q$ that hold throughout the entire domain and satisfy
\subeq{
  \begin{align}
    \rho \pd{u}{t} + \frac{1}{r^2}\pd{}{r}\left(r^2 q\right) &= 0, \qquad 0 < r < R_0, \label{eqn:energy} \\
    \tau \pd{q}{t} + q + k \pd{T}{r} &= 0, \qquad 0 < r < R_0, \label{eqn:mc} 
  \end{align}
}
where $\tau$ and $k$ represent the phase-dependent thermal relaxation time and conductivity, respectively.

% The standard heat equation can be obtained from \eqref{eqn:energy} using the thermodynamic relationship $\d u_l = c_l\, \d T_l$, which is valid for incompressible liquids (see Ref.~\cite[Chap.~2.3]{alexiades1993}), leading to
% \begin{align}
%   \rho_l c_l \pd{T_l}{t} + \frac{1}{r^2}\pd{}{r}\left(r^2q_l\right) = 0.
%   \label{eqn:heat}
% \end{align}

\subsection{Boundary conditions}

A no-flux condition is imposed on the solid flux at the origin:
\begin{align}
  q_s = 0, \quad r = 0.
\end{align}
The transfer of heat from the environment into the nanoparticle is modelled by a generalisation of Newton's law that accounts for the time scale of thermal relaxation. Relaxation effects can be incorporated into Newton's law by introducing a time derivative of the thermal flux \cite{cimmelli2010, jou2009, maurer1973}, leading to
\begin{align}
  \tau_l \pd{q_l}{t} + q_l = -h(T_e - T_l), \quad r = R_0,
  \label{eqn:newton}
\end{align}
where $h$ is a heat transfer coefficient. Assuming that the Maxwell--Cattaneo equation holds at the surface, Eqns.~\eqref{eqn:flux} and \eqref{eqn:newton} can be combined into a classical form of Newton's law given by
\begin{align}
-k_l \pd{T_l}{r} = -h(T_e - T_l), \quad r = R_0.
  \label{eqn:classical_newton}
\end{align}
It is important to note that the classical Newton boundary condition \eqref{eqn:classical_newton} only arises because of the specific forms of the non-classical Newton condition \eqref{eqn:newton} and the Maxwell--Cattaneo equation \eqref{eqn:flux}; it could not have been imposed \emph{a priori}. If an alternative model was used to describe non-Fourier heat conduction, then \eqref{eqn:classical_newton} would not be obtained.

An alternative and common choice for the boundary condition at  $r = R_0$ is to assume the temperature is held at a fixed value above the bulk melting temperature. However, as discussed by Ribera and Myers \cite{ribera2016}, the fixed-temperature condition leads to the melt front having an infinite initial velocity, which is not physically realistic. Newton-like boundary conditions avoid this behaviour by enabling the temperature at $r = R_0$ to gradually evolve from its initial condition rather than undergoing an instantaneous jump.

Conservation of energy across the solid-liquid interface can be enforced by applying the Rankine--Hugoniot condition to the energy balance \eqref{eqn:energy}, yielding a Stefan condition given by
% \begin{align}
%   \left[\rho_l L_m + \rho_l c_l (T_l - T_m^*) - \rho_s c_s(T_m - T_m^*)\right]\td{R}{t} = q_l, \quad r = R(t).
%   \label{eqn:stef_gen}
% \end{align}
\begin{align}
  \rho \left[L_m + c(T_l - T_s)\right]\dot{R} = q_l - q_s, \quad r = R(t),
  \label{eqn:stef_gen}
\end{align}
where $\dot{R} \equiv \d R / \d t$. In deriving \eqref{eqn:stef_gen} we have used the definitions of the internal energies given by \eqref{eqn:u} when evaluating the jump in internal energy across the interface.
%and the solutions for the solid temperature and flux stated in \eqref{eqn:solid}.

The final boundary conditions are the subject of some debate and depend on how the Maxwell--Cattaneo equation  is perceived. On one hand, the Maxwell--Cattaneo equation can be viewed as in \eqref{eqn:flux} as a constitutive relationship that is independently applied to the solid and liquid phase. In this case, it is natural to require the temperature to be continuous across the solid-liquid interface \cite{friedman1989, solomon1985}, leading to the boundary condition
\begin{align}
  T_s = T_l = T_m(R), \quad r = R(t),
  \label{eqn:T_cont}
\end{align}
where $T_m$ is the equilibrium melting temperature of a solid nanoparticle of radius $R$. 
The condition \eqref{eqn:T_cont} is obviously consistent with classical formulations of phase-change problems based on Fourier's law. The size-dependent melt temperature $T_m$ is given by the standard form of the Gibbs--Thomson condition, which is consistent with the assumption of constant specific heat capacity \cite{alexiades1993, font2013},
\begin{align}
  T_m(R) = T_m^*\left(1 - \frac{l_\text{cap}}{R}\right),
  \label{eqn:GT}
\end{align}
where $l_\text{cap}$ is the capillary length, \emph{i.e.}, the length scale on which surface energy becomes important. The capillary length can be written as $l_\text{cap} = (2 \sigma_{sl}) / (\rho_s L_m)$, where $\sigma_{sl}$ is the surface energy of the solid-liquid interface and $\rho_s$ is the density of the solid.

On the other hand, the Maxwell--Cattaneo equation can be viewed as in \eqref{eqn:mc} as a physical law that must hold throughout the two-phase mixture. In this case, it is natural to derive a jump condition from the Maxwell--Cattaneo equation \eqref{eqn:mc} in the same way that the Stefan condition \eqref{eqn:stef_gen} was derived from the local energy balance \eqref{eqn:energy}. If the thermal conductivity and relaxation time do not vary between the solid and liquid (\emph{i.e.,} if $\tau_s = \tau_l = \tau$ and $k_s = k_l = k$), then the Rankine--Hugoniot condition can be applied to \eqref{eqn:mc}, leading to
\begin{align}
  \tau (q_l-q_s) \td{R}{t} = k (T_l - T_s), \quad r = R(t).
  \label{eqn:mc_jump}
\end{align}
By combining \eqref{eqn:stef_gen} and \eqref{eqn:mc_jump}, a jump condition for the temperature \cite{greenberg1987, sobolev1991, sobolev1996} is obtained:
\begin{align}
  T_l - T_s = \frac{L_m}{c}\left(\frac{\kappa}{\tau \dot{R}^2} - 1\right)^{-1}, \quad r = R(t).
  \label{eqn:const_T_jump}
\end{align}
where $\kappa = k / (\rho c)$ is the thermal diffusivity. Naively taking $\tau \to 0$ in \eqref{eqn:const_T_jump} shows that temperature continuity is recovered in the classical limit, consistent with Fourier-based models of heat conduction. The limit as $\tau \to 0$ will be explored in greater detail in Sec.~\ref{sec:zero}.

The thermal conductivity of the solid phase is often greater than that of the liquid. A sharp jump (\emph{i.e.,} discontinuity) in thermal conductivity across the sharp solid-liquid interface prevents the Maxwell--Cattaneo equation \eqref{eqn:flux} from being put into conservative form and the Rankine--Hugoniot condition from being applied. An extended jump condition for the case of varying material properties can be derived by treating the solid-liquid interface as diffuse and performing a sharp-interface limit, details of which are presented in \ref{app:diffuse}. The extended jump condition can be written as
\begin{align}
  T_l(R,t) - T_s(R,t) =
  \frac{L_m}{2c} \left(\frac{\kappa_s}{\tau_s \dot{R}^2} - 1 \right)^{-1}
  +
  \frac{L_m}{2c} \left(\frac{\kappa_l}{\tau_l \dot{R}^2} - 1 \right)^{-1}.
  \label{eqn:T_jump}
\end{align}
The corresponding interfacial temperatures for the solid and liquid phases can also be determined from the sharp-interface limit and are found to be
\subeq{
  \label{eqn:T_inter}
  \begin{align}
    T_s(R,t) &= T_m(R) - \frac{L_m}{2c}\left(\frac{\kappa_s}{\tau_s \dot{R}^2}-1\right)^{-1}, \\
    T_l(R,t) &= T_m(R) + \frac{L_m}{2c}\left(\frac{\kappa_l}{\tau_l \dot{R}^2}-1\right)^{-1}.
  \end{align}
}
The interfacial conditions \eqref{eqn:T_inter} become singular when the melt front and thermal waves propagate with the same speed, $\dot{R} = (\kappa_i / \tau_i)^{1/2}$. In Sec.~\ref{sec:discussion}, it will be shown how the singular behaviour of the jump conditions prevent the onset of unphysical ``supersonic'' melting \cite{mullis1997, sobolev1991, sobolev2015ii}, whereby the speed of the melt front exceeds the speed of heat propagation.

Assuming that temperature continuity \eqref{eqn:T_cont} is true, then the Stefan condition \eqref{eqn:stef_gen} can be simplified to
\begin{align}
  \rho L_m \td{R}{t} = q_l - q_s, \quad r = R(t).
  \label{eqn:stef_cont}
\end{align}
When the jump condition based on \eqref{eqn:T_jump} is used in place of the continuity condition \eqref{eqn:T_cont}, the Stefan condition \eqref{eqn:stef_gen} becomes 
\begin{align}
  \rho L_m \left[1 + \frac{1}{2}\left(\frac{\kappa_l}{\tau_l \dot{R}^2}-1\right)^{-1} + \frac{1}{2}\left(\frac{\kappa_s}{\tau_s \dot{R}^2}-1\right)^{-1}\right]\dot{R} = q_l - q_s, \quad r = R(t).
\end{align}

\subsection{Initial conditions}

The initial temperature of the solid is assumed to be equal to the equilibrium melting temperature of a spherical nanoparticle of radius $R_0$. Furthermore, the initial flux through the solid is taken to be zero. Thus, we impose $T_s(r,0) = T_m(R_0)$ and $q_s(r,0) = 0$. Melting begins from the surface of the nanoparticle so that the initial position of the melt front is given by $R(0) = R_0$. As this corresponds to a nanoparticle that is completely solid when $t = 0$, it is not possible to impose initial conditions for the temperature and flux of the liquid. However, as shown in Sec.~\ref{app:smalltimes}, it is possible to obtain an asymptotic solution to the model that is valid for arbitrarily small times, which can be used in place of initial conditions.

\subsection{Non-dimensionalisation}
\label{sec:nondim}
The governing equations are now cast into dimensionless form by a suitable rescaling of the variables.  In doing so, we make the simplifying assumption that the model parameters are independent of temperature, although it is straightforward to account for their temperature dependence. Spatial variables are scaled by the initial radius of the solid nanoparticle $R_0$. We let $\Delta T = T_e - T_m(R_0)$ be the characteristic temperature scale of the system. As it is the influx of heat from the environment that drives the melting process, it is sensible to define a scale for the flux $q$ using the Newton boundary condition \eqref{eqn:newton}; this leads to $q \sim h \Delta T$. Finally, a time scale is chosen by balancing the change in latent heat with the thermal energy that is delivered to the interface, giving $t \sim (\rho L_m R_0) / (h \Delta T)$. Using these scales, we define dimensionless variables, marked by hats, given by $t = (\rho L_m R_0) / (h \Delta T) \hat{t}$, $r = R_0 \hat{r}$, $R(t) = R_0 \hat{R}(\hat{t})$, $T_i = T_m(R_0) + (\Delta T) \hat{T}_i$, and $q_i = (h \Delta T) \hat{q}_i$. The dimensionless bulk equations are given by (upon dropping the hats)
\subeq{
  \label{nd:bulk}
\begin{align}
  \beta^{-1} \pd{T_s}{t} + \frac{1}{r^2}\pd{}{r}\left(r^2 q_s \right) &= 0, \qquad 0 < r < R(t), \label{nd:ce_s} \\
  \gamma \tau_r \pd{q_s}{t} + q_s + k_r \mathcal{N}^{-1} \pd{T_s}{r} &= 0, \qquad 0 < r < R(t), \label{nd:mc_s} \\
  \beta^{-1} \pd{T_l}{t} + \frac{1}{r^2}\pd{}{r}\left(r^2 q_l \right) &= 0, \qquad R(t) < r < 1, \label{nd:ce_l} \\
  \gamma \pd{q_l}{t} + q_l + \mathcal{N}^{-1} \pd{T_l}{r} &= 0, \qquad R(t) < r < 1, \label{nd:mc_l}
\end{align}
}
where $\tau_r = \tau_s / \tau_l$, $k_r = k_s / k_l$, and $\gamma = \mathcal{C} \beta^{-1} \mathcal{N}$ is the effective relaxation parameter, which is the product of the Cattaneo number $\mathcal{C}$, the inverse Stefan number $\beta^{-1}$, and the Nusselt number $\mathcal{N}$.  These dimensionless numbers are defined as
\begin{align}
  %\gamma = \frac{\tau_r h \Delta T}{\rho_s L_m R_0}, \quad
  \mathcal{C} = \frac{\tau_l \kappa_l}{R_0^2}, \quad
  \beta = \frac{L_m}{c \Delta T}, \quad 
  \mathcal{N} = \frac{h R_0}{k_l}.
  \label{nd:numbers}
\end{align}
The Cattaneo number represents the ratio of the thermal relaxation time scale of the liquid to the classical thermal diffusion time scale.  The Stefan number characterises the time scale of melting relative to heat conduction.  Small values of the Stefan number are associated with fast melting processes.  The Nusselt number describes the rate at which thermal energy is transferred to the nanoparticle compared to the rate at which it is transported away from the surface by conduction. Large Nusselt numbers characterise situations involving rapid surface heating, with an infinite Nusselt number describing the case of an instantaneous jump in surface temperature. The effective relaxation parameter $\gamma$ therefore describes how the importance of thermal relaxation is affected by the time scales of thermal diffusion, melting, and heat transfer. 

The no-flux condition at the origin in dimensionless terms is given by
\begin{align}
  q_s = 0, \quad r = 0. \label{nd:no_flux}
\end{align}
The dimensionless Gibbs--Thomson condition for the melt temperature reads
\begin{align}\label{nd:GT}
  T_m(R) = \theta \ell \left(1 - \frac{1}{R}\right),
\end{align}
where 
$\theta = T_m^* / \Delta T$ is the dimensionless bulk melting temperature and $\ell = l_\text{cap} / R_0$ is the non-dimensional capillary length. 
% \begin{align}
%   \theta = \frac{l_\text{cap}}{R_0}\frac{T_m^*}{\Delta T}
%   %\theta = \frac{T_m^*}{\Delta T}, \quad \ell = \frac{l_\text{cap}}{R_0},
% \end{align}
% characterises the relative change in the melting temperature due to surface energy.
The cases of temperature continuity and a temperature jump can be combined into a single set of boundary conditions. The dimensionless interfacial conditions for the temperature are 
\subeq{
  \label{nd:T_bc}
  \begin{align}
    T_l(R,t) &= T_m(R) + \frac{\beta}{2}\left(\frac{\beta}{\mathcal{N} \gamma \dot{R}^2} - 1\right)^{-1},  \label{nd:T_bc_l} \\
    T_s(R,t) &= T_m(R) - \frac{\beta}{2}\left(\frac{k_r\beta}{\tau_r \mathcal{N} \gamma \dot{R}^2} - 1\right)^{-1}, \label{nd:T_bc_s}
  \end{align}
}
which lead to a Stefan condition given by
\begin{align}
  \left[1 + \frac{1}{2}\left(\frac{\beta}{\mathcal{N} \gamma \dot{R}^2}-1\right)^{-1} + \frac{1}{2}\left(\frac{k_r \beta }{\tau_r \mathcal{N}\gamma \dot{R}^2}-1\right)^{-1}\right]\dot{R} = q_l - q_s, \quad r = R(t).
  \label{nd:stef}
\end{align}
Taking the classical limit as $\gamma \to 0$ (with $\tau_r \gamma \to 0$) in \eqref{nd:T_bc}--\eqref{nd:stef}, with no additional rescaling of the variables, yields the boundary conditions for the case of a continuous temperature profile. The rescaled Newton boundary condition governing the influx of thermal energy is given by
\begin{align}\label{nd:newton}
  \pd{T_l}{r} = \mathcal{N}(1 - T_l), \quad r = 1.
\end{align}
Finally, the initial conditions for the solid temperature, solid flux, and position of the melt front are $T_s(r,0) \equiv 0$, $q_s(r,0) \equiv 0$, and $R(0) = 1$.

\subsection{Parameter estimation}
\label{sec:parameters}
We consider nanoparticles that have initial radii in the range of $10$~nm to $100$~nm. The lower bound of $10$~nm ensures that the melting process can be described using continuum theory, which is thought to hold on length scales greater than $2$~nm \cite{myers2014}. The material constants are based on tin, as this element is commonly used in experimental studies of nanoparticle melting \cite{bachels2000, jiang2006, lai1996}, making parameter values readily available.  However, we also provide parameter values for gold and lead in Table \ref{table:materials}.  
%Tin has a bulk melting temperature of $T_m^* = 505$~K and a latent heat of melting given by $L_m = 58.5$~kJ/kg. The density of solid and liquid tin is $\rho_s = 7180$~kg/m\unit{3} and $\rho_l = 6980$~kg/m\unit{3}, respectively. The heat capacities are $c_s = 230$~J/(kg$\cdot$K) and $c_l = 268$~J/(kg$\cdot$K). The thermal conductivity of liquid and solid tin is $k_l = 30$~W/(m$\cdot$K) and $k_l = 67$~W/(m$\cdot$K). 
Following Bachels \etal\cite{bachels2000}, we take the surface energy of the solid-liquid interface to be $\sigma_{sl} = 0.055$~J/m\unit{2}.  The corresponding capillary length is  $l_\text{cap} = 0.26$~nm.  Estimates of the thermal relaxation time of metallic systems typically range from $10^{-12}$~s to $10^{-10}$~s \cite{mullis1997}.  The heat transfer coefficient $h$ is difficult to estimate due to its dependence on the environment that surrounds the nanoparticle. However, as discussed by Ribera and Myers \cite{ribera2016}, it is possible to define a theoretical maximum value for the heat transfer coefficient, $h_\text{max}$, that is consistent with the laws of thermodynamics and independent of the environment surrounding the nanoparticle. Ribera and Myers state that $h_\text{max} = 4.7 \times 10^{9}$~W/(m\unit{2}$\cdot$K) for tin. The largeness of $h_\text{max}$ reflects the fact that it provides the closest approximation to the fixed-temperature boundary condition that is possible without instantly vapourising the nanoparticle. We restrict our attention to the case where $h = h_\text{max}$ and leave exploring the melting behaviour for smaller values of the heat transfer coefficient as an area of future work.

The ratios of the specific heat capacities and densities of tin are $c_s / c_l = 0.86$ and $\rho_s / \rho_l = 1.03$, justifying our assumption that these remain constant during melting. The ratio of the bulk thermal conductivities is $k_r = 2.23$.  The Stefan number is given by $\beta = (235\,\text{K})/\Delta T$ and will typically be large unless the temperature difference $\Delta T$ is on the order of 100~K.  The dimensionless bulk melting temperature $\theta = (505\,\text{K}) / \Delta T$ will have the same order of magnitude as the Stefan number.  The dimensionless capillary lengths are small and range from $\ell = 2.6\E{-3}$ to $2.6\E{-2}$ as $R_0$ decreases from 100~nm to 10~nm.  The Nusselt number can be expressed in terms of the initial radius as $\mathcal{N} = (0.157\,\text{nm}^{-1})\,R_0$ and will range from $\mathcal{N}_\text{max}$ = 1.57 to 15.7.  Rather than estimate the order of magnitude of the Cattaneo number $\mathcal{C}$, it is more enlightening to work directly with the effective relaxation parameter, which may be written as $\gamma = (\tau_l h \Delta T) / (\rho L_m R_0)$.  Using the maximum heat transfer coefficient $h_\text{max}$, $\tau_l = 10^{-10}$~s, and $R_0 = 10$~nm, the effective relaxation parameter can be written in terms of the temperature difference as $\gamma = (0.114\,\text{K}^{-1})\Delta T$.  A value of $\Delta T \simeq 10$~K is reasonable and would make the effective relaxation parameter $O(1)$ in size.  However, for tin nanoparticles that exceed 10~nm in radius or which have reduced heat transfer coefficients or thermal relaxation times, the effective relaxation parameter will be smaller.

\begin{table}
    \caption{Thermophysical parameter values for some materials \cite{ribera2016,font2013}.}
    \label{table:materials}
\centering
	\begin{tabular}{|c|c|c|c|c|c|c|}
    	\hline
		Material & $T_m^*$ [K] & $L_m^*$ [kJ/kg] & $\rho_s/\rho_l$ [kg/m\unit{3}] & $c_s/c_l$ [J/(kg$\cdot$K)] & $k_s/k_l$ [W/(m$\cdot$K)] & $\sigma_{sl}$ [J/m\unit{2}] \\ \hline
        Tin & 505 & 58.5 & 7180/6980 & 230/268 & 67/30 & 0.055 \\ \hline
        Gold & 1337 & 63.7 & 19300/17300 & 129/163 & 317/106 & 0.27 \\ \hline
        Lead & 600 & 23.0 & 11300/10700 & 128/148 & 35/16 & 0.05 \\ \hline
        %Aluminium & 660 & 321.0 & 2550/2380 &  &  & 0.093--0.14  \\ \hline
		%Silver & 1234.95 & 111.0 &  &  &  &  \\ \hline
	\end{tabular}
\end{table}

%-----------------------------------------------------

\section{Numerical method}\label{sec:numerics}
The non-dimensional model \eqref{nd:bulk}--\eqref{nd:newton} is numerically solved using a semi-implicit finite difference method on a staggered grid \cite{rieth2018}. The transformations $\xi = r / R(t)$ and $\eta = (r - R(t)) / (1 - R(t))$ are used to map the evolving domains $0 < r < R(t)$ and $R(t) < r < 1$ to stationary domains $0 < \eta,\xi < 1$, respectively. The stationary domains are then discretised into cells of uniform size. Second-order central differences are used to approximate spatial derivatives. The temperature is calculated at cell edges while the flux is calculated at cell midpoints. By staggering the grid in this manner, the usual second-order finite-difference scheme for the heat equation is recovered in the limit of zero relaxation time. Linear interpolation and extrapolation are used to map quantities from one grid to another and to evaluate the flux at points that lie outside of the physical domain. Given known values of the position and speed of the melt front, $R$ and $\d R / \d t$, the temperature and the flux are implicitly solved using a backwards Euler iteration. With the newly computed values for the flux and temperature, the Stefan condition \eqref{nd:stef} is then numerically solved to produce a new value of $\d R / \d t$. The position of the melt front $R$ is then updated using a forward Euler iteration. This approach leads to a decoupling of the melting and transport processes and avoids the need to solve a nonlinear system of equations at each time step.

%-----------------------------------------------------

\section{Asymptotic analysis}\label{sec:asymptotics}
Asymptotic methods are used to calculate analytical solutions to the governing equations under specific limits and simplifying assumptions.  We first focus on calculating an approximate solution that is valid for arbitrarily small times. This solution can be used as a consistent initial condition when numerically simulating the model.  Furthermore, the small-time solution gives key insights into the physical relevance of the jump condition for the temperature.  We then examine the ``classical'' limit as $\gamma \to 0$, showing that the Fourier solution is only recovered in the case of a continuous temperature profile.  Finally, we construct solutions that are valid in the limit of large Stefan number. Large Stefan numbers characterise phase-change processes that are slow compared to the rate of thermal diffusion. As this is the case in many physical scenarios and for a range of materials, the large-Stefan number limit is perhaps the most common approach for simplifying models of phase change. The relatively fast rate of thermal diffusion implies that on the time scale of melting, the temperature can be well approximated by its quasi-static profile. An analytical solution for the temperature can then often be obtained, which allows the problem to be reduced to a system of ordinary differential equations for position of the free boundary and, in the case of non-Fourier heat conduction \cite{hennessy2018}, the thermal flux.  

%-----------------------------------------------------

\subsection{Solution for small times}\label{app:smalltimes}
The behaviour of the solution for small times can be determined by introducing an artificial small parameter $\varepsilon \ll 1$ and then writing $t = \varepsilon \bar{t}$. It will also be convenient to use the Maxwell--Cattaneo equation \eqref{nd:mc_l} to write the Newton condition \eqref{nd:newton} as
\begin{align}
  \gamma \varepsilon^{-1} \pd{q_l}{\bar{t}} + q_l = -(1 - T_l), \quad r = 1.
  \label{st:newton}
\end{align}
Separate balancing procedures are used to determine the relevant scales for variables associated with the liquid and solid phases.  In the case of the liquid, the scales are chosen to ensure the initial influx of thermal energy from the environment is captured by the model. In the case of the solid, the scales are chosen to capture the development of a thermal boundary layer near the melt front, which forms due to the change in melting temperature with nanoparticle size. To facilitate the calculation, we will assume that all of the dimensionless parameters (aside from $\varepsilon$) are $O(1)$ in magnitude. 

\subsubsection{Fourier conduction}
Setting the effective relaxation parameter $\gamma$ to zero recovers the classical formulation of the melting problem where the flux is governed by Fourier's law. The bulk equations can be written as
\subeq{
  \label{st:bulk}
\begin{alignat}{2}
  \beta^{-1} \pd{T_s}{\bar{t}} &= \frac{\varepsilon k_r \mathcal{N}^{-1}}{r^2}\pd{}{r}\left(r^2 \pd{T_s}{r} \right), &\qquad &0 < r < R(t),
  \label{st:T_s} \\
  \beta^{-1} \pd{T_l}{\bar{t}} &= \frac{\varepsilon \mathcal{N}^{-1}}{r^2}\pd{}{r}\left(r^2 \pd{T_l}{r} \right), &\qquad &R(t) < r < 1, \label{st:T_l}
\end{alignat}
}
and the Stefan condition is
\begin{align}
  \varepsilon^{-1} \td{R}{\bar{t}} = k_r\mathcal{N}^{-1}\pd{T_s}{r} - \mathcal{N}^{-1}\pd{T_l}{r}, \quad
  r = R(\bar{t}). \label{st:stef_classical}
\end{align}
To determine the problem for the liquid, we start by balancing the left- and right-hand sides of the Newton condition \eqref{nd:newton} that the temperature gradient $\pdf{T_l}{r}$ must be $O(1)$ in size. Using these facts in the Stefan condition \eqref{nd:stef} leads to the conclusion that $1 - R = O(\varepsilon)$. Therefore, we write $R = 1 - \varepsilon \bar{R}$. Using this rescaling for $R$ in the Gibbs--Thomson condition \eqref{nd:GT} then indicates that $T_m = O(\varepsilon)$ as $\varepsilon \to 0$. Therefore, we rescale the variables according to $r = 1 - \varepsilon \bar{r}$, $T_l = \varepsilon \bar{T}_l$, and $q_l = \bar{q}_l$. After taking the limit $\varepsilon\to0$, the heat equation \eqref{st:T_l} becomes
% \begin{equation}, and, and
% 	  \pd{^2\bar{T}}{\bar{r}^2}= 0, \label{st:eq_fourier}
% \end{equation}
% subject to the boundary conditions
% \subeq{
% \begin{alignat}{2}
%   \pd{\bar{T}}{\bar{r}} &= -\mathcal{N}, &\quad \bar{r} &= 0, \label{st:newton_cont_fourier}\\
%   \bar{T} &= -\theta\ell \bar{R}, &\quad \bar{r} &= \bar{R}(\bar{t}). \label{st:GT_fourier}
% \end{alignat}
% }
$\partial^2 \bar{T}_l / \partial \bar{r}^2 = 0$, which is subject to the boundary conditions $\pdf{\bar{T}_l}{\bar{r}} = -\mathcal{N}$ at $\bar{r} = 0$ and $\bar{T}_l = -\theta \ell \bar{R}$ at $\bar{r} = \bar{R}(\bar{t})$.  Upon solving for the liquid temperature, the Stefan condition \eqref{st:stef_classical} reduces to
\begin{align}\label{st:stefan_fourier}
  \td{\bar{R}}{\bar{t}}=1 + k_r \mathcal{N}^{-1}\pd{T_s}{r}, \quad r = R(t). 
\end{align}
All that remains is to determine the problem for the solid. The evolution of the solid temperature is driven by changes in the melting temperature. Since $T_m = O(\varepsilon)$, we scale $T_s = \varepsilon \bar{T}$. Naively taking $\varepsilon \to 0$ in \eqref{st:T_s} and using the initial condition shows that $\bar{T}_s(r,\bar{t}) \equiv 0$, which cannot satisfy the boundary condition at the melt front $\bar{T}_s(1,t) = -\theta \ell \bar{R} + O(\varepsilon)$. To capture the formation of a boundary layer at the melt front, we use the change of variable $r = 1 - \varepsilon \bar{R}(\bar{t}) - \varepsilon^{1/2} \bar{\zeta}$ and $\bar{t} = \bar{t}'$, which leads to the problem for the solid temperature given by (after dropping the prime)
\begin{align}
  \beta^{-1}\pd{\bar{T}}{\bar{t}} = k_r \mathcal{N}^{-1}\pdd{\bar{T}_s}{\bar{\zeta}} + O(\varepsilon^{1/2}),
\end{align}
with $\bar{T}_s = -\theta \ell \bar{R}$ at $\bar{\zeta} = 0$, $\bar{T}_s \to 0$ as $\bar{\zeta} \to \infty$, and $\bar{T}_s = 0$ at $\bar{t} = 0$. Using the scales for the solid problem in the Stefan condition \eqref{st:stefan_fourier} shows that
\begin{align}\label{st:stefan_fourier_2}
  \td{\bar{R}}{\bar{t}}=1 - \varepsilon^{1/2} k_r \mathcal{N}^{-1}\left.\pd{\bar{T}_s}{\bar{\zeta}}\right|_{\bar{\zeta} = 0},
\end{align}
implying that the thermal conduction through the solid does not strongly influence the initial melting kinetics. Thus, to leading order, $\bar{R}({\bar{t}}) = \bar{t}$. In principle, the solution for the melt front can be used in the problem for the solid temperature, which can be solved using a similarity solution of the form $\bar{T}_s(\bar{\zeta}, \bar{t}) = \bar{t} f(\bar{\zeta}/\bar{t}^{1/2})$; however, we do not compute this here.

In terms of the original non-dimensional variables, the small-time solutions for the liquid temperature, liquid flux, and position of the melt front are given by
\subeq{
\label{st:fourier_soln}
\begin{align}
T_l(r,t) &= \mathcal{N}[r - R(t)] - \theta \ell [1 - R(t)], \\
q_l(r,t) &= -1, \label{st:fourier_q} \\
R(t) &= 1 -  t. \label{st:fourier_R}
\end{align}
}
Importantly, \eqref{st:fourier_q} shows that the liquid flux predicted by Fourier's law is $O(1)$ in size and constant in time, which results in the linear melting kinetics given by \eqref{st:fourier_R}.  The magnitude of the flux ultimately arises from the form of the Newton condition \eqref{st:newton}.  Since thermal relaxation effects have been neglected, the surface flux, and hence the bulk flux, instantaneously attain $O(1)$ values.  As we will show in Sec.~\ref{sec:st_cont} and Sec.~\ref{sec:st_jump}, this will not be the case when relaxation effects are included in the model.  The small-time solution given by \eqref{st:fourier_soln} is equivalent to that found in other studies of nanoparticle melting \cite{ribera2016} and phase change \cite{mitchell2009}. Furthermore, the linear kinetics given by \eqref{st:fourier_R} have also been observed in models of drug diffusion through polymeric spheres \cite{mccue2011}.

% For the remainder of the section, we will focus on obtaining small-time solutions for positive $O(1)$ values of the effective relaxation parameter $\gamma$.  We will therefore replace Fourier's law with the Maxwell--Cattaneo equation given by \eqref{nd:mc} and use the full form of the Newton boundary condition \eqref{st:newton}.

\subsubsection{Maxwell--Cattaneo conduction with temperature continuity}
\label{sec:st_cont}
We follow the same procedure to determine the scales for the solid and liquid problems in the non-classical melting problem. Balancing terms in the Newton boundary condition \eqref{st:newton} requires the liquid flux to be $O(\varepsilon)$ in size, $q_l = \varepsilon \bar{q}_l$.  From the Stefan condition \eqref{nd:stef}, the small magnitude of the liquid flux implies that the width of the melt region is $O(\varepsilon^2)$. Although conduction through the solid could influence the early melting kinetics, this will be shown not to be the case. Thus, we let $R = 1 - \varepsilon^2 \bar{R}$. The Gibbs--Thomson condition \eqref{nd:GT} then indicates that the temperatures can be written as $T_l = \varepsilon^2 \bar{T}_l$ and $T_s = \varepsilon^2 \bar{T}_s$.

To capture the thermal boundary layer in the solid problem, we let $r = 1 - \varepsilon^2 \bar{R}(\bar{t}) - \varepsilon \bar{\zeta}$. Balancing thermal relaxation and temperature gradients in the Maxwell--Cattaneo equation \eqref{nd:mc_s} leads to a flux scale of $O(\varepsilon^2)$. Thus, we write $q_s = \varepsilon^2 \bar{q}_s$, showing that the solid flux is indeed smaller than the liquid flux. As $\varepsilon \to 0$, the leading-order problem for the solid temperature and flux is given by
\subeq{
  \label{st:T_s_cont}
\begin{align}
  \beta^{-1} \pd{\bar{T}_s}{\bar{t}} - \pd{\bar{q}_s}{\bar{\zeta}} &= 0, \\
  \gamma \tau_r \pd{\bar{q}_s}{\bar{t}} - k_r \mathcal{N}^{-1} \pd{\bar{T}_s}{\bar{\zeta}} &= 0,
\end{align}
}
with $\bar{T}_s = -\theta \ell \bar{R}$ at $\bar{\zeta} = 0$, $\bar{T}_s \to 0$ as $\bar{\zeta} \to \infty$, and $\bar{T}_s = \bar{q}_s = 0$ at $\bar{t} = 0$. Combining both equations in \eqref{st:T_s_cont} shows that the temperature satisfies a wave equation.

The problem for the liquid can be formulated by writing $r = 1 - \varepsilon^2 \bar{r}$. Taking $\varepsilon \to 0$ results in a significantly reduced model. Conservation of energy \eqref{nd:ce_l} becomes $\pdf{\bar{q}_l}{\bar{r}} = 0$, implying the flux is only a function of time.  The Maxwell--Cattaneo equation \eqref{nd:mc_l} reduces to
%\subeq{
%  \label{st:bulk_c}
%\begin{align}
%  \pd{\bar{q}}{\bar{r}} &= 0, \label{st:q}\\
%  \gamma\, \pd{\bar{q}}{\bar{t}} &= \mathcal{N}^{-1} \pd{\bar{T}}{\bar{r}}.\label{st:T}
%\end{align}
%}
\begin{align}
  \gamma\, \td{\bar{q}_l}{\bar{t}} &= \mathcal{N}^{-1} \pd{\bar{T_l}}{\bar{r}}.\label{st:T}
\end{align}
The boundary conditions read
\subeq{
\begin{alignat}{2}
  \gamma\, \td{\bar{q}_l}{\bar{t}} &= -1, &\quad \bar{r} &= 0, \label{st:newton_cont}\\
  \bar{T}_l &= -\theta\ell \bar{R}, &\quad \bar{r} &= \bar{R}(\bar{t}). \label{st:GT}
\end{alignat}
}
We assume that there is no flux through the surface of the nanoparticle at time $\bar{t} = 0$.
The leading-order Stefan condition is
\begin{align}
  \td{\bar{R}}{\bar{t}} = -\bar{q}_l,
  \quad
  \bar{r} = \bar{R}(\bar{t}).
  \label{st:stefan}
\end{align}
The solution for the liquid temperature and flux, as well as the melt front, written in terms of the original dimensionless variables, is given by
\subeq{
  \label{st:soln_cont_l}
  \begin{align}
    T_l(r,t) &= \mathcal{N}[r - R(t)] - \theta \ell [1 - R(t)], \\
    q_l(r,t) &= -\gamma^{-1} t, \label{st:soln_cont_q} \\
    R(t) &= 1 - \frac{1}{2}\,\gamma^{-1} t^2.\label{R_smalltimes}
  \end{align}
}
By accounting for thermal relaxation effects, the magnitude of the liquid flux \eqref{st:soln_cont_q} is now predicted to increase linearly with time.  This is in stark contrast to the instantaneous change in the liquid flux that is observed when using Fourier's law to describe heat conduction.  The gradual change of the liquid flux in the case of Maxwell--Cattaneo conduction leads to substantially slower melting kinetics, which are now quadratic in time.

\subsubsection{Maxwell--Cattaneo conduction with temperature jump}
\label{sec:st_jump}

At leading order, the small-time solutions for the cases of temperature continuity and a temperature jump are the same. This is because the initially slow rate of melting, $\dot{R} = O(t)$, leads to a temperature jump that is  $O(t^2)$ in magnitude; see Eqn \eqref{nd:T_bc}. Thus, the jump does not appear in the leading-order problem for the small-time solution.

\subsubsection{Summary of small-time behaviours}

  The small-time analysis performed here, in combination with those carried out by other authors, have identified four scaling laws for the initial melting kinetics. When using a fixed-temperature boundary condition, Fourier's law, and neglecting density variation between phases (see, \emph{e.g.} Refs.~\cite{mccue2009, back2014, font2013}), the width of the liquid region has been shown to grow like $t^{1/2}$. When density variation is included \cite{font2015}, this scaling law changes to $t^{3/4}$. Both of these scaling laws lead to infinite growth rates as $t \to 0$. This is because the fixed-temperature boundary condition results in an infinite temperature gradient and thus to an infinite thermal flux. When a Newton condition is used, the width of the liquid region scales like $t$, avoiding an infinite initial growth rate. This behaviour is explained through the fact the temperature at $r = 1$ gradually decreases from zero, so the temperature gradient remains finite and the flux through the liquid is given by $q_l \simeq -1$.  In the case of Fourier conduction, this scaling law holds regardless of whether the density variation is considered \cite{ribera2016}. When thermal relaxation is included in the Newton boundary condition, the scaling law changes to $t^{2}$, reflecting the additional time that is required for the system to respond to the change in thermal environment, leading to $q_l \simeq -t$.

\subsection{Limit of zero thermal relaxation time}\label{sec:zero}
We now examine the ``classical'' limit given by $\gamma \to 0$ with all other parameters $O(1)$ in size.  We first assume that the temperature is continuous across the melt front and then consider the case of a temperature jump. 

\subsubsection{Temperature continuity}
\label{sec:z_tc}
There are two key time regimes in the limit as $\gamma \to 0$. The first time regime, $t = O(\gamma)$, captures the quick thermal relaxation that occurs as thermal energy is being transferred to the nanoparticle.  In the second time regime, $t = O(1)$, the flux has relaxed to its Fourier value and the classical nanoparticle melting problem is recovered. 

The scales for the first time regime are determined using the same balancing procedure that was used to construct small-time solutions. The pertinent scales for the solid problem are given by $t = \gamma \tilde{t}$, $R = 1 - \gamma \tilde{R}$, $T_s = \gamma \tilde{T}_s$, and $q_s = \gamma^{1/2} \tilde{q}_s$. The thermal boundary layer has width $O(\gamma^{1/2})$. The solid problem is analogous to that in \eqref{st:T_s_cont} but involves the full Maxwell--Cattaneo equation. The relevant scales for the liquid problem are $r = 1 - \gamma \tilde{r}$, $T_l = \gamma \tilde{T}_l$, and $q_l = \tilde{q}_l$. After taking $\gamma \to 0$, the energy equation \eqref{nd:ce_l} reduces to $\pdf{\tilde{q}_l}{\tilde{r}} = 0$, implying the flux is constant in space.  The Maxwell--Cattaneo equation \eqref{nd:mc_l} becomes
\begin{align}
  \td{\tilde{q}_l}{\tilde{t}} + \tilde{q}_l &= \mathcal{N}^{-1}\pd{\tilde{T}_l}{\tilde{r}}.
  \label{z:mc_j}
\end{align}
Differentiating \eqref{z:mc_j} with respect to $r$ shows that the temperature satisfies Laplace's equation.  The boundary conditions for the temperature are given by $\pdf{\tilde{T}_l}{\tilde{r}} = -\mathcal{N}$ at $\tilde{r} = 0$ and $\tilde{T}_l = -\theta \ell \tilde{R}$ at $\tilde{r} = \tilde{R}(\tilde{t})$.  Upon solving for the temperature and inserting the result into \eqref{z:mc_j}, we find that the flux and the position of the melt front satisfy the coupled system of equations
\subeq{
\begin{align}
  \td{\tilde{q}_l}{\tilde{t}} + \tilde{q}_l &= -1, \\
  \td{\tilde{R}}{\tilde{t}} &= -\tilde{q}_l,
  \end{align}
  }
subject to $\tilde{q}_l = \tilde{R} = 0$ when $\tilde{t} = 0$. The solution for the liquid temperature, liquid flux, and position of the melt front, written in terms of the original dimensionless variables, is given by
% \begin{alignat}{2}
%   \pd{\tilde{T}}{\tilde{r}} &= -\mathcal{N}, &\quad \tilde{r} &= 0, \\
%   \tilde{T} &= -\theta \ell \tilde{R}, &\quad \tilde{r} &= \tilde{R}(\tilde{t}).
% \end{alignat}
% The Stefan condition 
%In terms of the original dimensionless variables, we find that
\subeq{
\label{z:temp_cont_soln}
\begin{align}
T_l(r,t) &= \mathcal{N}[r - R(t)] - \theta \ell [1 - R(t)], \\
q_l(r,t) &= -[1 - \exp(-t / \gamma)], \label{z:q_c} \\
R(t) &= 1 - \left[t + \gamma \exp(-t/\gamma) - \gamma\right]. \label{z:R_c}
\end{align}
}
As $t \to 0$ with $\gamma$ fixed, the solutions for the flux \eqref{z:q_c} and the position of the melt front \eqref{z:R_c} reduce to
\begin{align}
q_l(r,t) \sim -\gamma^{-1}t , \quad R(t) \sim 1 - \frac{1}{2}\gamma^{-1}t^2,
\end{align}
in agreement with the small-time limit of the non-Fourier model given in \eqref{st:soln_cont_l}.  

In the second time regime, defined by $t = O(1)$, the relaxation terms can be neglected and the variables do not need to be rescaled from their original dimensionless form.  The Maxwell--Cattaneo equations \eqref{nd:mc_s} and \eqref{nd:mc_l} therefore reduce to Fourier's law.  The matching conditions for the flux and the position of the melt front can be obtained from \eqref{z:q_c} and \eqref{z:R_c} in the limit $\gamma \to 0$ with $t$ fixed, yielding
\begin{align}
q_l(r,t) \sim -1, \quad R(t) \sim 1 -  t,
\label{z:matching_cont}
\end{align}
which are precisely the small-time solutions to the Fourier model \eqref{st:fourier_soln}.  Consequently, non-classical transport mechanisms do not enter the leading-order problem on $O(1)$ time scales, implying that the classical, Fourier-based model is completely recovered when temperature continuity is imposed at the melt front.

\subsubsection{Temperature jump}

There are again two time regimes to consider, $t = O(\gamma)$ and $t = O(1)$. However, the asymptotic solutions in these regimes are the same as when the temperature is continuous across the melt front. The scaling in the first regime implies that $\dot{R} = O(1)$. From the interfacial conditions in \eqref{nd:T_bc}, the temperature jump is $O(\gamma)$ and thus does not influence the leading-order solution when $\gamma \to 0$. The same argument applies in the second time regime as well. Thus, the classical, Fourier-based solution can also be recovered from the limit $\gamma \to 0$ when the temperature jump is used with the Maxwell--Cattaneo equation.

\subsection{Solution for large Stefan numbers}
We now compute asymptotic solutions in the limit as $\beta \to \infty$ with $\gamma = O(1)$.
%in the case of a continuous temperature profile.  We do not repeat the calculation for the case of a temperature jump due the unphysical nature of the model with this boundary condition.  More specifically, for all of the materials listed in Table \ref{table:materials}, the assumption of a large Stefan number leads to the (dimensional) external number $T_e$ being smaller than the effective melting temperature $(k_s / k_l) T_m^*$.  Therefore, melting is simply not possible in the case of a large Stefan number and a temperature jump at the melt front. 
Before proceeding with the calculations, it is helpful to re-examine the sizes of the various non-dimensional numbers that appear in the model.  The parameter $\theta$ is of the same order of magnitude as the Stefan number, implying that $\theta = O(\beta)$.  The dimensionless capillary length $\ell$ is always small.  We will make an additional assumption that $\theta \ell = O(1)$ as $\beta \to \infty$ so that the temperature at the melt front is $O(1)$ during the majority of the melting process.  For practical scenarios, where $\Delta T$ is of the order of a few Kelvin or larger, this assumption will be true.

\subsubsection{Maxwell--Cattaneo conduction with temperature continuity}
\label{sec:large_beta_cont}

There are two main time regimes to consider. The first regime is defined by $t = O(\beta^{-1/2})$ and captures non-Fourier heat conduction through the solid core of the nanoparticle. Only a small amount of melting occurs during the first time regime. The second time regime, defined by $t = O(1)$, captures the complete melting of the solid core. The temperature profiles are quasi-steady in the second time regime, but non-Fourier transport mechanisms still affect the liquid flux, altering the melting kinetics.

The first time regime is captured by writing $t = \beta^{-1/2} \hat{t}$, $R = 1 - \beta^{-1} \hat{R}$, $q_s = \beta^{-3/2} \hat{q}_s$, $q_l = \beta^{-1/2} \hat{q}_l$, $T_s = \beta^{-1} \hat{T}_s$ and $T_l = \beta^{-1} \hat{T}_l$. The problem for the solid temperature is given by
\subeq{\label{lb:solid1}
\begin{align}
  \pd{\hat{T}_s}{\hat{t}} + \frac{1}{r^2}\pd{}{r}\left(r^2 \hat{q}_s\right) &= 0, \label{lb:solid_energy1}\\
  \gamma \tau_r \pd{\hat{q}_s}{\hat{t}} + \beta^{-1/2} \hat{q}_s + k_r \mathcal{N}^{-1}\pd{\hat{T}_s}{r} &= 0, \label{lb:solid_mc1}
\end{align}
}
with $\hat{q}_s = 0$ at $r = 0$, $\hat{T}_s = -\theta \ell \hat{R} + O(\beta^{-1})$ at $r = 1 + O(\beta^{-1})$, and $\hat{T}_s = \hat{q}_s = 0$ when $\bar{t} = 0$. The problem for the liquid and the melt front can be obtained by letting $r = 1 - \beta^{-1}\hat{r}$ to find
\subeq{
  \begin{align}
    \pd{\hat{q}_l}{\hat{r}} + O(\beta^{-1}) &= 0, \\
    \gamma \pd{\hat{q}_l}{\hat{t}} + \beta^{-1/2}\hat{q}_l - \mathcal{N}^{-1}\pd{\hat{T}_l}{\hat{r}} &= 0,
                                                                                                       \label{lb:liquid_mc1}
  \end{align}
}
with $\pdf{\hat{T}_l}{\hat{r}} = -\mathcal{N} + O(\beta^{-1})$ at $\hat{r} = 0$ and $\hat{T}_l = -\theta \ell \hat{R} + O(\beta^{-1})$ at $\hat{r} = \hat{R}(\hat{t})$.  The Stefan condition is $\d \hat{R}/\d \hat{t} = -\hat{q}_l + O(\beta^{-1})$.

The leading-order solution for the liquid flux and temperature is given by
$\hat{T}_l = -\mathcal{N}(\hat{r} - \hat{R}) - \theta \ell \hat{R}$ and $\hat{q}_l = -\hat{t}$, where $\hat{R} = -(1/2)\gamma^{-1}\hat{t}^2$. A closed-form solution to the leading-order solid problem is not required. However, it is interesting to examine the large-time behaviour in order to understand how the second time regime is entered. The Gibbs--Thomson condition implies that the solid temperature behaves like $\hat{T}_s = O(\hat{t}^2)$ as $\hat{t} \to \infty$. The scaling for the flux is determined by balancing both terms in the energy equation \eqref{lb:solid_energy1}, yielding $\hat{q}_s = O(\hat{t})$. Using these scales in the Maxwell--Cattaneo equation \eqref{lb:solid_mc1} reveals that the temperature gradient dominates all other terms. Thus, we find that
\begin{align}
  \hat{T}_s \sim -\frac{1}{2}\,\theta \ell \gamma^{-1}\hat{t}^2, \quad \hat{q}_s \sim \frac{1}{3}\,\theta \ell \gamma^{-1}\hat{t} r \label{lb:s_sol1}
\end{align}
as $\hat{t} \to \infty$. The time scale for the second regime is obtained from the liquid problem. Using the leading-order solutions for the liquid temperature and flux in \eqref{lb:liquid_mc1} shows that all three terms balance when $\hat{t} = O(\beta^{1/2})$, corresponding to $t = O(1)$.

To study the dynamics in the second time regime, where $t = O(1)$, we first note that $T_s = O(1)$ and $q_s = O(\beta^{-1})$ due to the limiting behaviour in the first time regime given by \eqref{lb:s_sol1}. Additionally, $r = O(1)$ in the solid region, implying that the leading-order solution for the solid problem is given by
  \begin{align}
    T_s = T_m(R), \quad q_s = \frac{1}{3}\,\beta^{-1} r \td{T_m}{t}.
  \end{align}
  We also find that $q_l = O(1)$ in the second regime, implying that the solid will not influence the melting kinetics.
% Thus, taking $\beta \to \infty$ in \eqref{nd:ce_s} shows that the solid flux is uniform in space and equal to zero due to the no-flux condition at the origin \eqref{nd:no_flux}. By using this solution in \eqref{nd:mc_s}, it follows that the solid temperature is also uniform and equal to the melting temperature. Thus, we have that $T_s(r,t) = T_m(R)$ and $q_s(r,t) = 0$, the latter of which implies the solid phase does not influence the melting kinetics.}
The leading-order contributions to \eqref{nd:ce_l} show that the liquid flux $q_l$ is divergence free and can be written as
\begin{equation}\label{red_s:q}
	q_l(r,t)=\frac{Q_l(t)}{r^2},
\end{equation}
where $Q_l(t)$ is the volume-averaged flux defined as
\begin{align}
Q_l(t) = \frac{1}{1 - R(t)}\int_{R(t)}^{1}q_l(r,t)r^2\,\d r.
\label{eqn:Q}
\end{align}
To calculate the temperature profile for the liquid, we apply the divergence operator to the Maxwell--Cattaneo equation \eqref{nd:mc_l} and use the fact the liquid fluxes are divergence free, which shows the liquid temperature satisfies Laplace's equation. Although this is similar to the case of Fourier conduction, we emphasise that the liquid flux still satisfies the full Maxwell--Cattaneo equation \eqref{nd:mc_l}.
%By inserting \eqref{red_s:q} into the Maxwell--Cattaneo equation \eqref{nd:mc_l}, multiplying by $r^2$, and then differentiating with respect to $r$, the temperature is found to .
After solving Laplace's equation and imposing temperature continuity and the Newton condition \eqref{nd:newton}, the temperature profile is given by
\begin{align}
T_l(r,t) = \theta \ell \left[1 - \frac{1}{R(t)}\right] + \frac{\mathcal{N}\big\{R(t) + \theta\ell [1 - R(t)]\big\}}{R(t) + \mathcal{N}[1 - R(t)]}\left[\frac{1}{R(t)} - \frac{1}{r}\right].
\label{red_s:T}
\end{align}
Upon substituting \eqref{red_s:q} and \eqref{red_s:T} into \eqref{nd:mc_l}, a differential equation for the mean flux $Q_l$ is obtained,
\subeq{
\label{red_s:odes}
\begin{equation}\label{ode_A_largebeta}
	\gamma \td{Q_l}{t} + Q_l =-\frac{R+\theta\ell\left(1 - R\right)}{R+\mathcal{N}(1-R)}.
\end{equation}
By substituting the solution the liquid flux into the Stefan condition \eqref{nd:stef} and neglecting the small contribution from the solid flux, we find that the position of the melt front evolves according to
\begin{equation}\label{ode_R_largebeta}
  \td{R}{t}=\frac{Q_l}{R^2}.
  %\td{R}{t}=\frac{Q}{R^2}.
\end{equation}
}
%Although the second term on the left-hand side of \eqref{ode_R_largebeta} can generally be neglected, it does not lead to additional complications and thus it is retained.
An initial condition for $Q_l$ can be obtained via a small-time analysis. Following the methodology of Sec.~\ref{app:smalltimes}, we find that the small-time behaviour of $Q_l$ and $R$ is given by
\begin{align}
  Q_l(t) \sim -\gamma^{-1}t, 
  \quad 
  %R(t) \sim 1 - \frac{1}{2}\,\gamma^{-1} t^2,
  R(t) \sim 1 - \frac{1}{2}\gamma^{-1} t^2,
  \label{red_s:ic}
\end{align}
as $t \sim 0$. The limiting behaviour in \eqref{red_s:ic} exactly coincides with the small-time behaviour of the full model, implying that the solution to \eqref{red_s:odes} will be uniformly valid for all times. The equations in \eqref{red_s:odes} along with \eqref{red_s:ic} form a reduced model that can be used to easily study the melting dynamics of nanoparticles.

\subsubsection{Maxwell--Cattaneo conduction with temperature jump}

We now suppose that the temperature has a jump at the melt front. The interfacial conditions for the temperature \eqref{nd:T_bc} can be written in the limit $\beta \to \infty$ as
  \subeq{
    \begin{align}
      T_l(R,t) &= T_m(R) + \frac{1}{2}\mathcal{N}\gamma \dot{R}^2 + O(\beta^{-1}), \label{lb:Tl}\\
      T_s(R,t) &= T_m(R) - \frac{1}{2} k_r^{-1} \mathcal{N}\gamma \tau_r \dot{R}^2 + O(\beta^{-1}).
    \end{align}
  }
  The analysis follows in the same manner as in Sec.~\ref{sec:large_beta_cont} for a continuous temperature profile. The scalings and solutions in the first time regime, $t = O(\beta^{-1/2})$, remain unchanged. This is because $\dot{R} = O(\beta^{-1/2})$ in the first time regime and thus the temperature jump does not appear in the leading-order problem.  Differences do occur in the second time regime, $t = O(1)$, however. The solid flux is still small, $q_s = O(\beta^{-1})$, and the solid temperature is uniform, $T_s(r,t) = T_s(R(t),t)$. The liquid temperature satisfies Laplace's equation and the solution can be written as
  \begin{align}
    T_l(r,t) = \frac{\mathcal{N}(1 - T_l(R,t))}{1 + \mathcal{N}(R^{-1} - 1)}\left(\frac{1}{R} - \frac{1}{r}\right) + T_l(R,t),
  \end{align}
  where $T_l(R,t)$ is given by \eqref{lb:Tl}. The reduced system of equations governing the melting process can be written as
  \subeq{
    \label{red:ode_jump}
    \begin{align}
      \gamma \td{Q_l}{t} + Q_l &= -\frac{R + \theta \ell(1 - R)}{R + \mathcal{N}(1 - R)} + \frac{\gamma}{2}\frac{\mathcal{N}}{R + \mathcal{N}(1 - R)}\frac{Q_l^2}{R^3}, \\
      \td{R}{t} &= \frac{Q_l}{R^2}.
    \end{align}
  }
  The small-time behaviour is given by \eqref{red_s:ic}, which can be used as initial conditions when solving \eqref{red:ode_jump} numerically.

\section{Results and discussion}\label{sec:discussion}

The asymptotic analysis of Sec.~\ref{sec:asymptotics} reveals that the Maxwell--Cattaneo equation can lead to marked changes in the melting kinetics compared to Fourier's law.  Furthermore, for sufficiently small times, the choice of boundary condition does not alter the predicted melting kinetics.  To explore how the choice of boundary condition affects the long-term melting behaviour, we carry out numerical simulations of the full governing equations. In the following discussion, we will refer to the MC-TC and MC-TJ models as those based on the Maxwell--Cattaneo equation with temperature continuity and a temperature jump at the melt front, respectively.

In the first numerical experiment, we compare predictions obtained using all three models: Fourier, MC-TC, and MC-TJ.  We consider a nanoparticle of initial radius $R_0 = 10$~nm and a Stefan number of $\beta = 10$, corresponding to $\Delta T \simeq 24$~K.  We suppose the solid and liquid have the same relaxation time and set $\tau_s = \tau_l = 10^{-10}$~s, leading to $\tau_r = 1$ and $\gamma \simeq 2.5$.

%The asymptotic analysis of Sec.~\ref{sec:zero} reveals that when the Maxwell--Cattaneo equation is used to model heat conduction, the predicted melting behaviour is strongly dependent on the choice of boundary condition applied at the melt front. This is true even when the effective relaxation parameter is small. To explore the difference in melting dynamics in further detail, we numerically solve the governing equations using both boundary conditions and a small value of the relaxation parameter. Carrying out simulations using both boundary conditions for the same parameter values requires the Stefan number to be small: the jump condition may only be imposed when the temperature difference is sufficiently (and unrealistically) large. Thus, we set $R_0 = 10$~nm, $\tau_r = 10^{-13}$~s, and $\beta = 0.25$, corresponding to $\Delta T \simeq 873$~K, which leads to $\gamma \simeq 0.10$. 
% The chosen value of $\beta$ corresponds to an imposed temperature difference of $\Delta T \simeq 873$~K, which is large but necessary to ensure that melting occurs when the jump condition for the temperature is imposed. The small value of the relaxation time is used to offset the factor of $\beta^{-1}$ that appears in the definition of $\gamma$. 

\begin{figure}
  \centering
    \centering
    \subfigure[]{\includegraphics[width=0.49\textwidth]{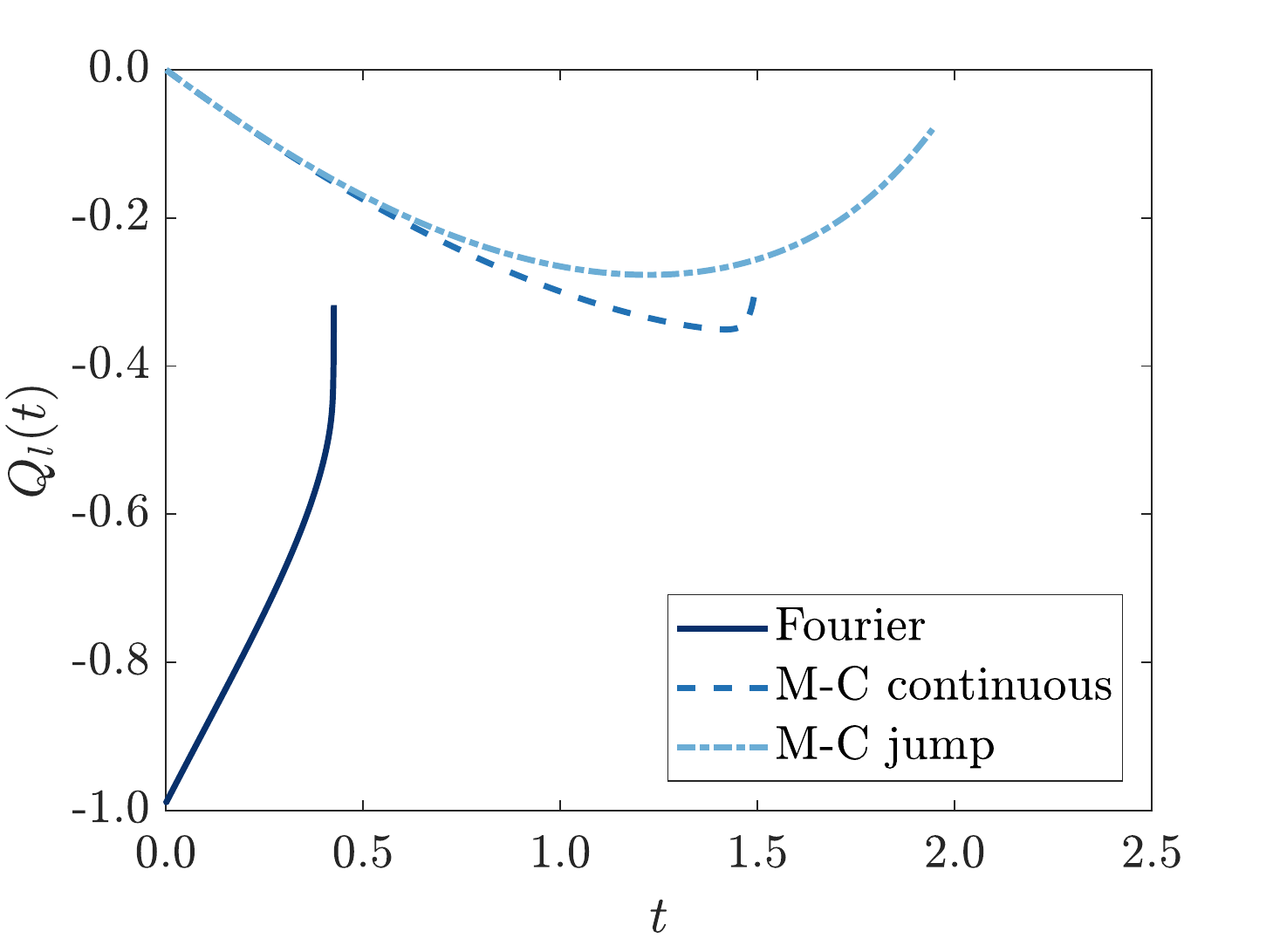}}
    \subfigure[]{\includegraphics[width=0.49\textwidth]{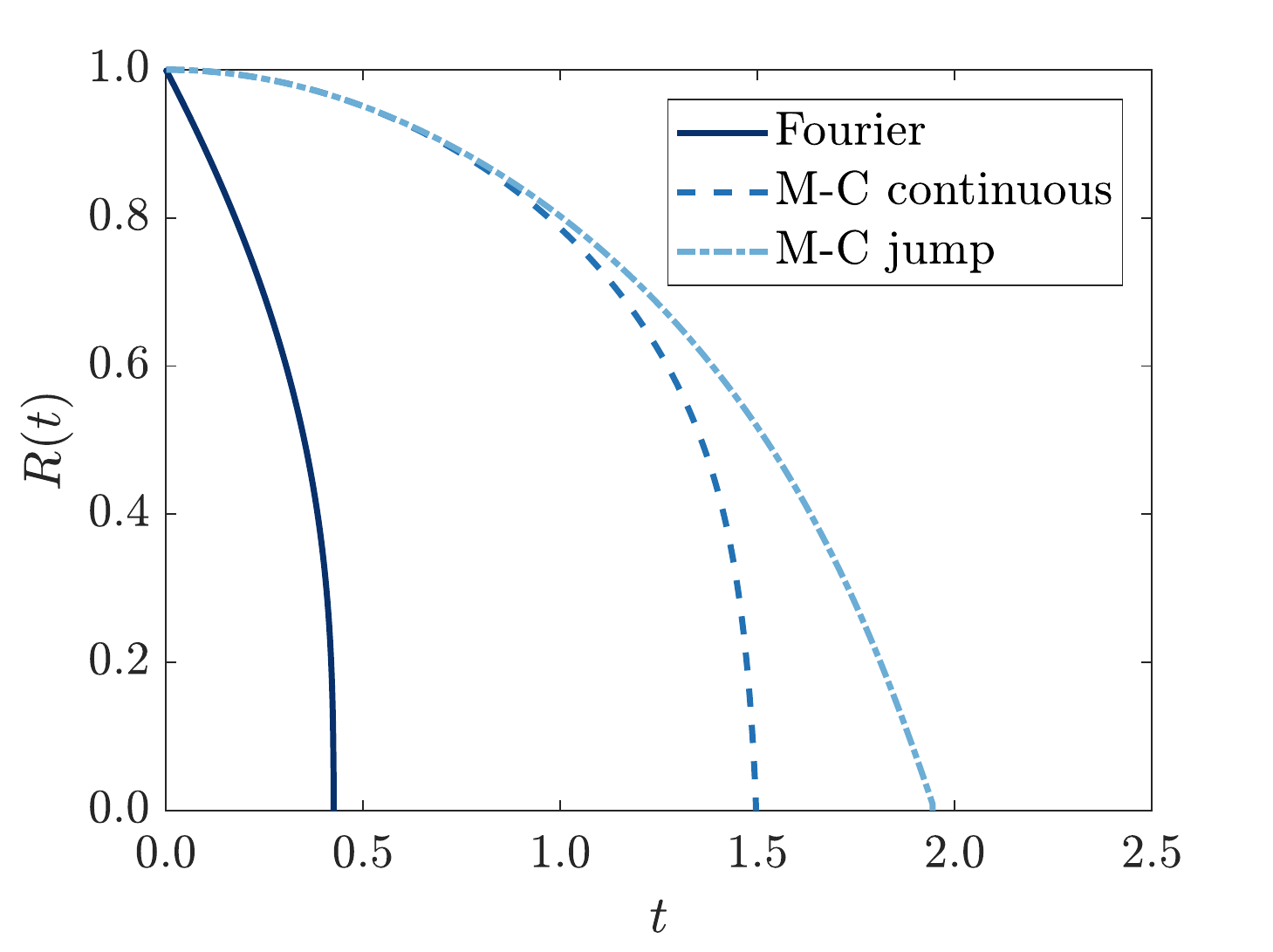}}
    \caption{Comparison of melting dynamics obtained using Fourier's law (solid lines), the Maxwell--Cattaneo equation with temperature continuity at the melt front (dashed lines), and the Maxwell--Cattaneo equation with a temperature jump at the melt front (dashed-dotted lines). 
      (a) Evolution of the mean liquid flux defined by \eqref{eqn:Q}. 
      (b) Evolution of the position of the melt front. 
      Parameter values are $R_0 = 10$~nm, $\tau_s = \tau_l = 10^{-10}$~s and $\beta = 10$, corresponding to $\gamma \simeq 2.5$ and $\Delta T \simeq 24$~K.}
  \label{fig:all_comp}
\end{figure}

\begin{figure}
  \centering
  \centering
  \subfigure[]{\includegraphics[width=0.49\textwidth]{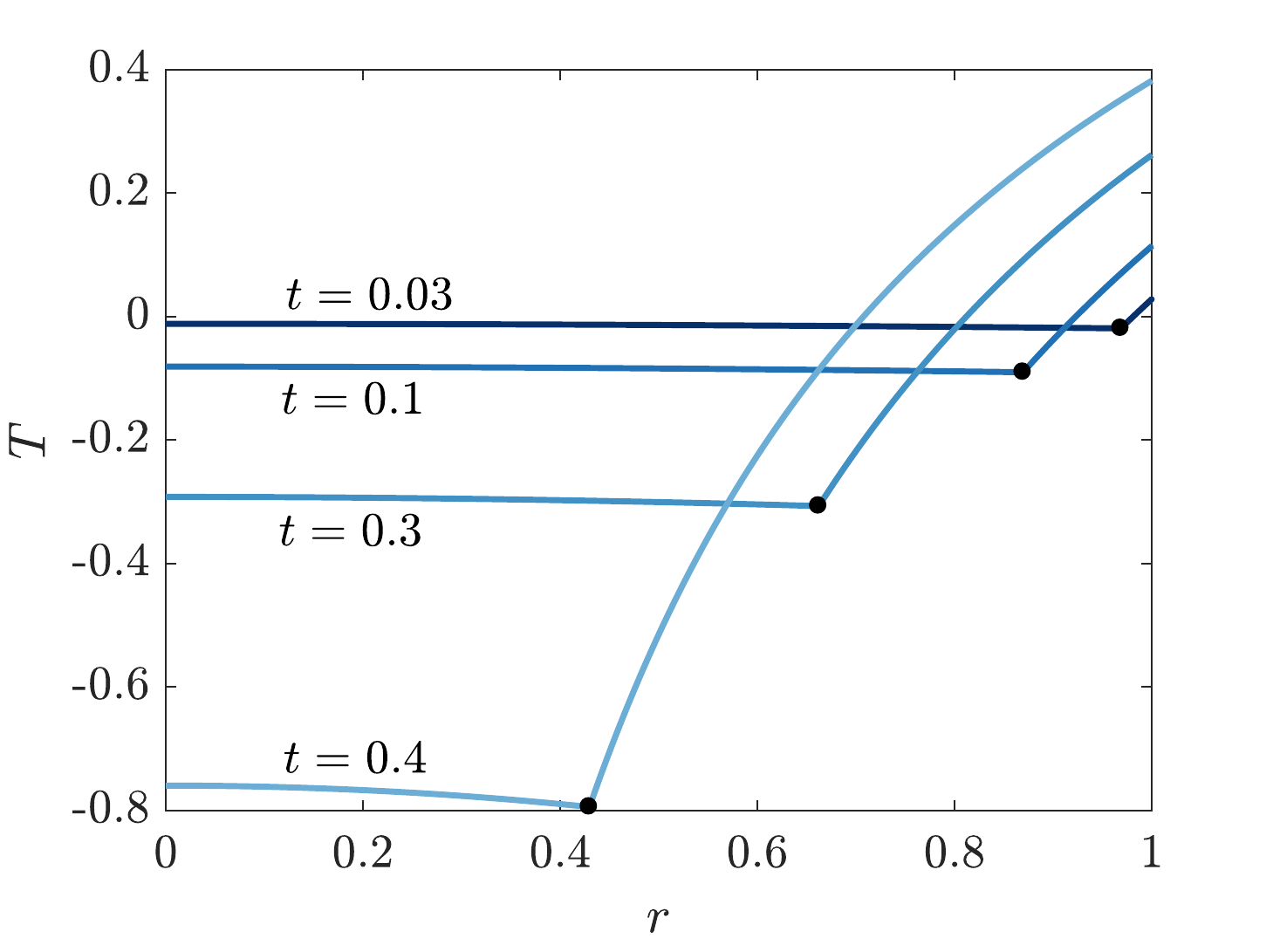}}
    \subfigure[]{\includegraphics[width=0.49\textwidth]{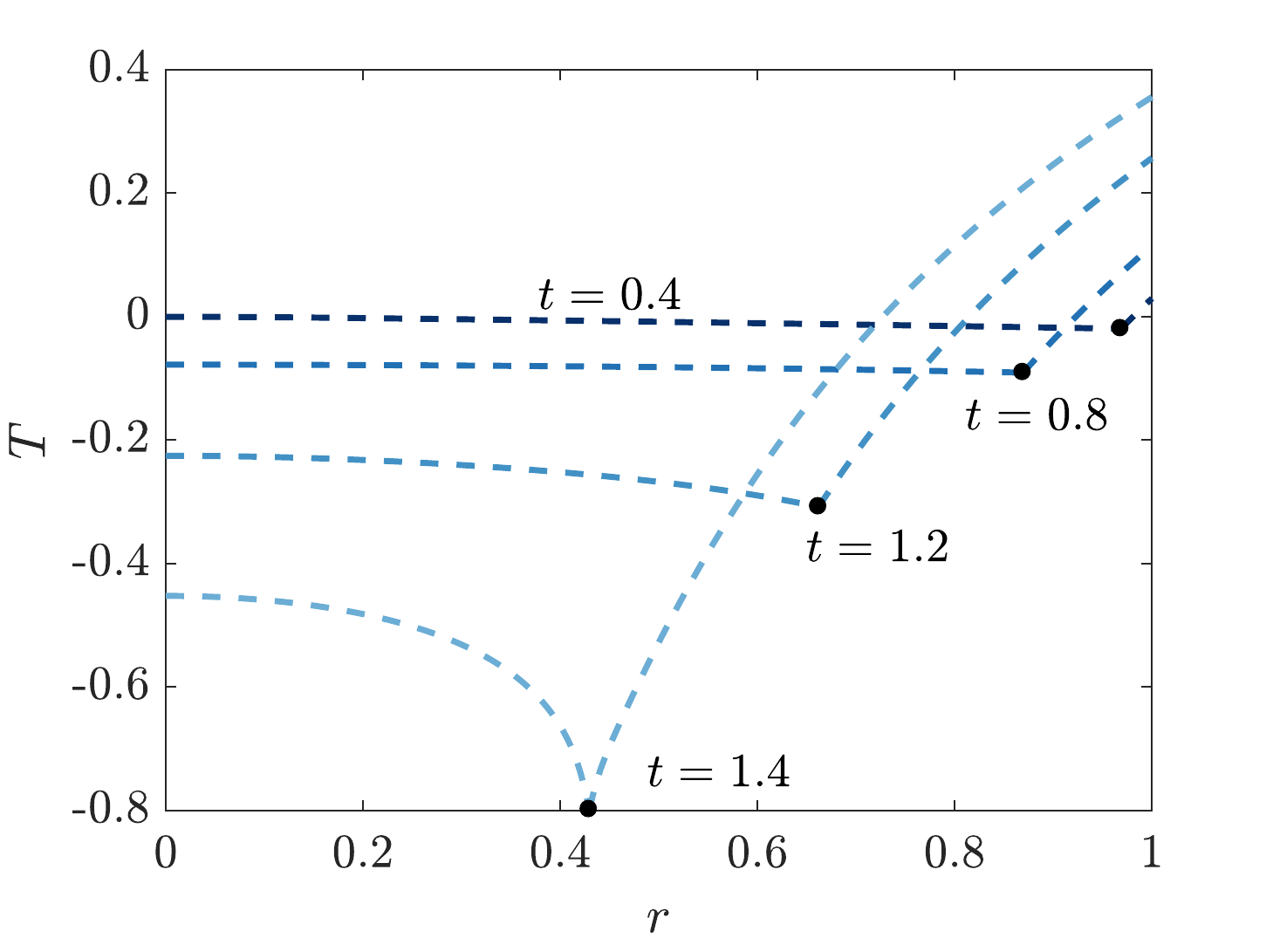}}
    \subfigure[]{\includegraphics[width=0.49\textwidth]{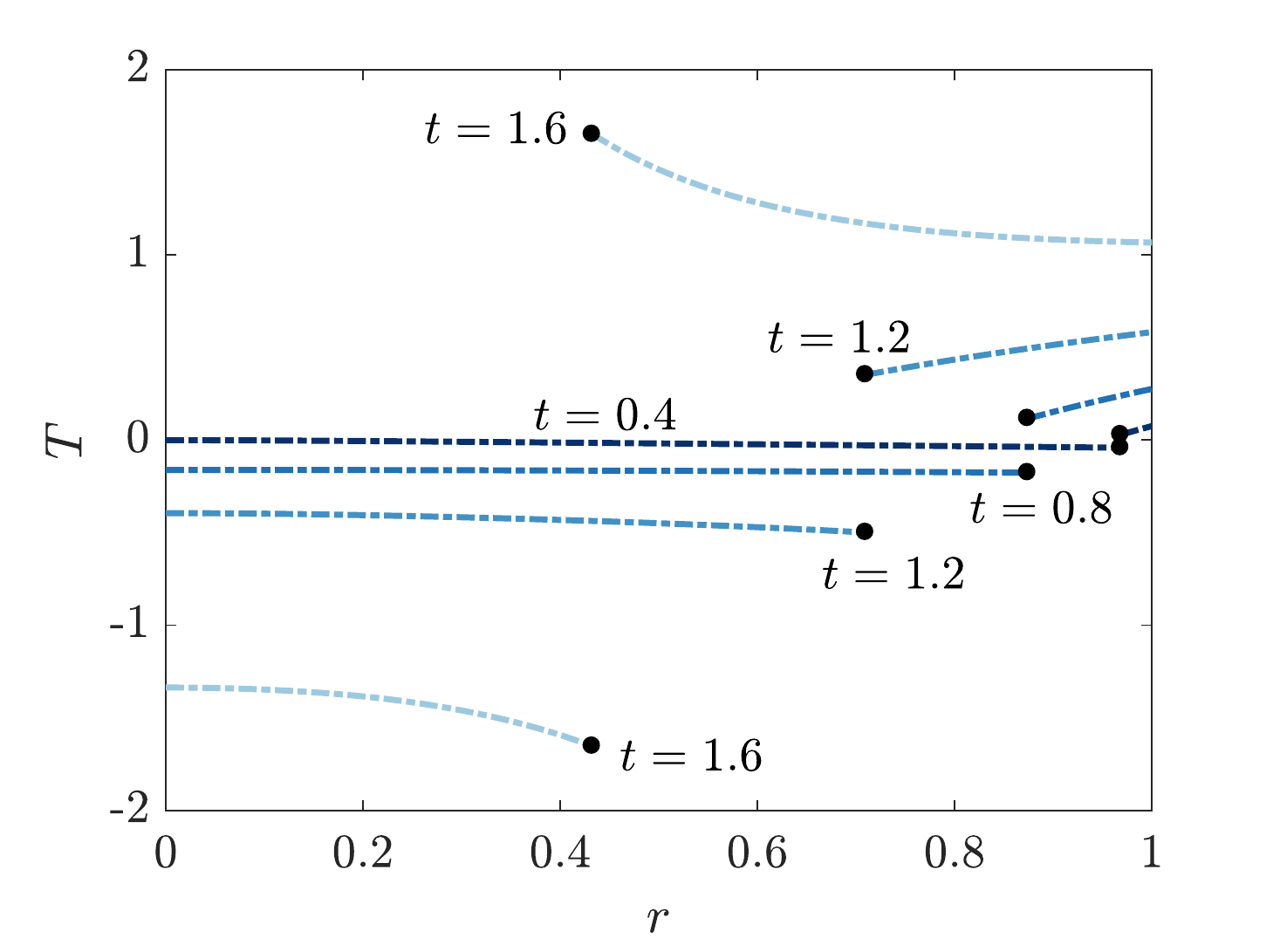}}
    \caption{Comparison of temperature profiles obtained using Fourier's law (a), the Maxwell--Cattaneo equation with temperature continuity at the melt front (b), and the Maxwell--Cattaneo equation with a temperature jump at the melt front (c). The solid dot denotes the position of the solid-liquid interface.  Parameter values are the same as those in Fig.~\ref{fig:all_comp}: $R_0 = 10$~nm, $\tau_s = \tau_l = 10^{-10}$~s and $\beta = 10$, corresponding to $\gamma \simeq 2.5$ and $\Delta T \simeq 24$~K.}
  \label{fig:temp}
\end{figure}

The results of the simulations are shown in Fig.~\ref{fig:all_comp}, which plots the evolution of the mean liquid flux $Q_l(t)$ defined in \eqref{eqn:Q} and the position of the melt front $R(t)$. The temperature profiles are shown in Fig.~\ref{fig:temp}.  
%The circles and diamonds in Fig.~\ref{fig:all_comp} denote the asymptotic predictions of the flux and position of the melt front given by \eqref{z:temp_cont_soln} and \eqref{z:tj_sol_ogamma}, respectively, which accurately capture the evolution of the system on $O(\gamma)$ time scales. 
As predicted by the asymptotic analysis, Fourier's law (solid line) leads to the mean thermal flux instantaneously obtaining the value of $Q_l = -1$ and then monotonically decreasing in magnitude. When the Maxwell--Cattaneo equation is used (dashed and dashed-dotted lines), the magnitude of the flux grows in time from its initial value of zero until it reaches a maximum and begins to decay.  The location of the maximum marks the transition to the diffusion-dominated transport regime.  For this set of parameter values, the Maxwell--Cattaneo equation leads to a substantially slower melting process compared to Fourier's law.  In the case of temperature continuity (dashed lines), the entire melting process takes place in the non-classical regime, whereby heat conduction is dominated by relaxation effects.  In the case of a temperature jump across the melt front (dashed-dotted lines), the melting process is slightly slower compared to the case of temperature continuity, and the diffusion-dominated transport regime is entered.

The different melting behaviour that is observed when using the Maxwell--Cattaneo equation but varying the boundary condition can be rationalised in terms of the temperature at the melt front (see Fig.~\ref{fig:temp}). Imposing the jump condition leads to the temperature of the liquid at the melt front being larger than the Gibbs--Thomson melting temperature. Consequently, there is a smaller temperature difference between the liquid and the environment and the transfer of heat into the nanoparticle is reduced. This means that less thermal energy is transported to the melt front, resulting in a slower melting process compared to when the temperature profile is assumed to be continuous.  

Figures \ref{fig:temp} (a) and (b) show the temperature profiles obtained from the Fourier and MC-TC models at four times that correspond to the same positions of the melt front. The parameter values are $R_0 = 10$~nm,  $\beta = 10$, and $\tau_s = \tau_l = 10^{-10}$~s and correspond to those of Fig.~\ref{fig:all_comp}.  During the early stages of the melting process, when $R > 0.6$, the temperature profiles in the two models are quantitatively similar.  However, as the solid core shrinks and the melting process is accelerated, the MC-TC model predicts that extremely large gradients in the solid temperature form near the melt front whereas the Fourier model predicts a relatively uniform temperature. We attribute this difference in behaviour to the fact that the MC-TC model captures the finite thermal response time of the solid, which becomes increasingly relevant during the final stages of melting. In particular, the bulk temperature is unable to respond to the rapidly changing melting temperature and, therefore, a large thermal gradient develops at the melt front. In the Fourier and MC-TC models, the speed of the melt front increases in time and becomes unbounded as the radius of the solid core approaches zero. This leads to a ``thermal sonic boom'' in the MC-TC model, whereby the speed of the melt front equals and then exceeds the finite speed of thermal waves. In dimensionless terms, the speed of heat propagation in the solid and liquid is given by $v_s = \sqrt{k_r \beta / (\tau_r \gamma \mathcal{N})}$ and $v_l = \sqrt{\beta / (\gamma \mathcal{N})}$, respectively. Since $v_s > v_l$ for these parameter values, the transition from ``subsonic melting'' with $\dot{R} < v_l$ to ``supersonic melting'' with $\dot{R} > v_l$ occurs at roughly $t = 1.4$. After this time, we find that the temperature profile in the liquid develops a local maximum near the melt front, a feature which is impossible in Fourier-based heat conduction. To explain this phenomenon, we recall that in the supersonic regime, the melt front propagates faster than thermal energy can be delivered to it from the liquid. The system therefore responds by introducing a local maximum in the temperature, amplifying the local temperature gradient and flux, thus compensating for the insufficient rate of thermal transport from the liquid. Ultimately, supersonic melting is unphysical because the speed of melting cannot exceed the speed of heat propagation \cite{sobolev1991} and, as discussed below, the onset of a local maximum in the temperature profile violates the second law of thermodynamics.  It is well known \cite{bright2009} that solutions obtained from the Maxwell--Cattaneo equation can violate the second law and thus we do not attribute the observed behaviour to numerical errors.

  The temperature profiles obtained from the MC-TJ model are shown in Fig.~\ref{fig:temp} (c). As the solid core shrinks and the speed of the melt front increasing, the temperature jump at the interface grows in magnitude. Eventually, the temperature of the liquid at the melt front exceeds that at the outer boundary, leading to heat flowing in the direction of positive temperature gradient, in violation of the second law. The increase in liquid temperature at the melt front reduces the temperature gradient across the liquid and thus the thermal flux, slowing the melting process and thereby preventing the onset of supersonic melting. Thus, the jump condition for the temperature essentially acts as a flux limiter, forcing the temperature profiles to adjust in such as way to limit the flux and avoid the onset of supersonic phase change.

%The results of our analysis highlight two key shortcomings of the jump condition for the temperature. Firstly, unrealistically large temperature differences (or external temperatures) are required to initiate melting. Secondly, it is not possible to recover the classical, Fourier-based solution in the limit as the effective relaxation parameter tends to zero. These two shortcomings ultimately stem from the large change in interfacial temperature that occurs as a result of the thermal conductivity of the solid being greater than that of the liquid. In previous studies that model phase change with the Maxwell--Cattaneo equation and use the temperature jump condition \cite{greenberg1987, glass1991, sobolev1996}, the thermal conductivities of the liquid and solid are assumed to be equal, thus avoiding the shortcomings that are observed here. Greenberg \cite{greenberg1987} even states that it is the jump condition, not the continuity condition, that leads to solutions that are in agreement and consistent with those obtained using Fourier's law. However, our asymptotic and numerical analyses show that this statement is simply not true.  We believe that our results provide sufficient evidence to invalidate the jump condition.  Thus, we will exclusively focus on the case of a continuous temperature profile throughout the remainder of this section. 

\begin{figure}
  \centering
  \subfigure[Temperature continuity]{\includegraphics[width=0.49\textwidth]{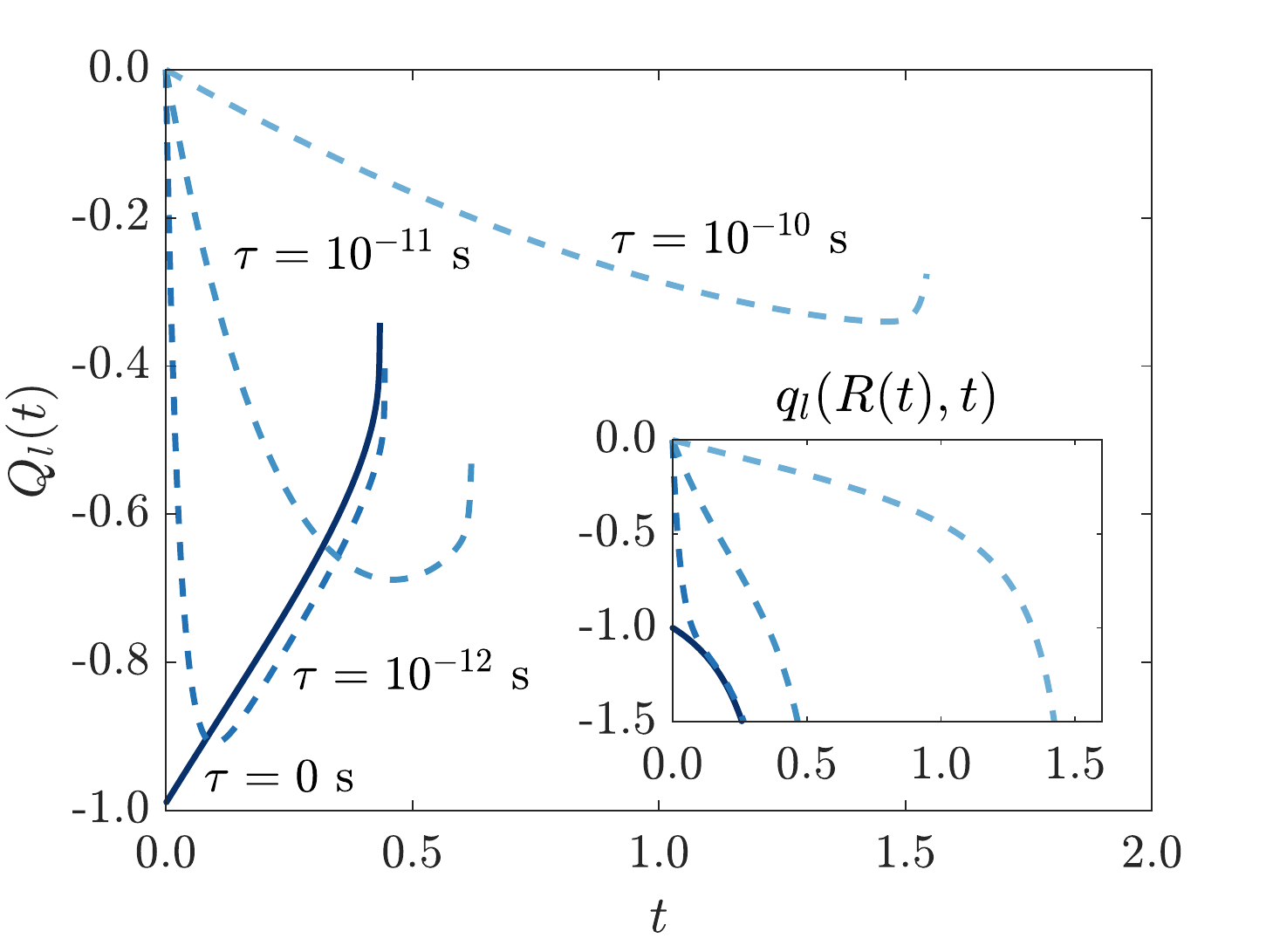}}
  \subfigure[Temperature continuity]{\includegraphics[width=0.49\textwidth]{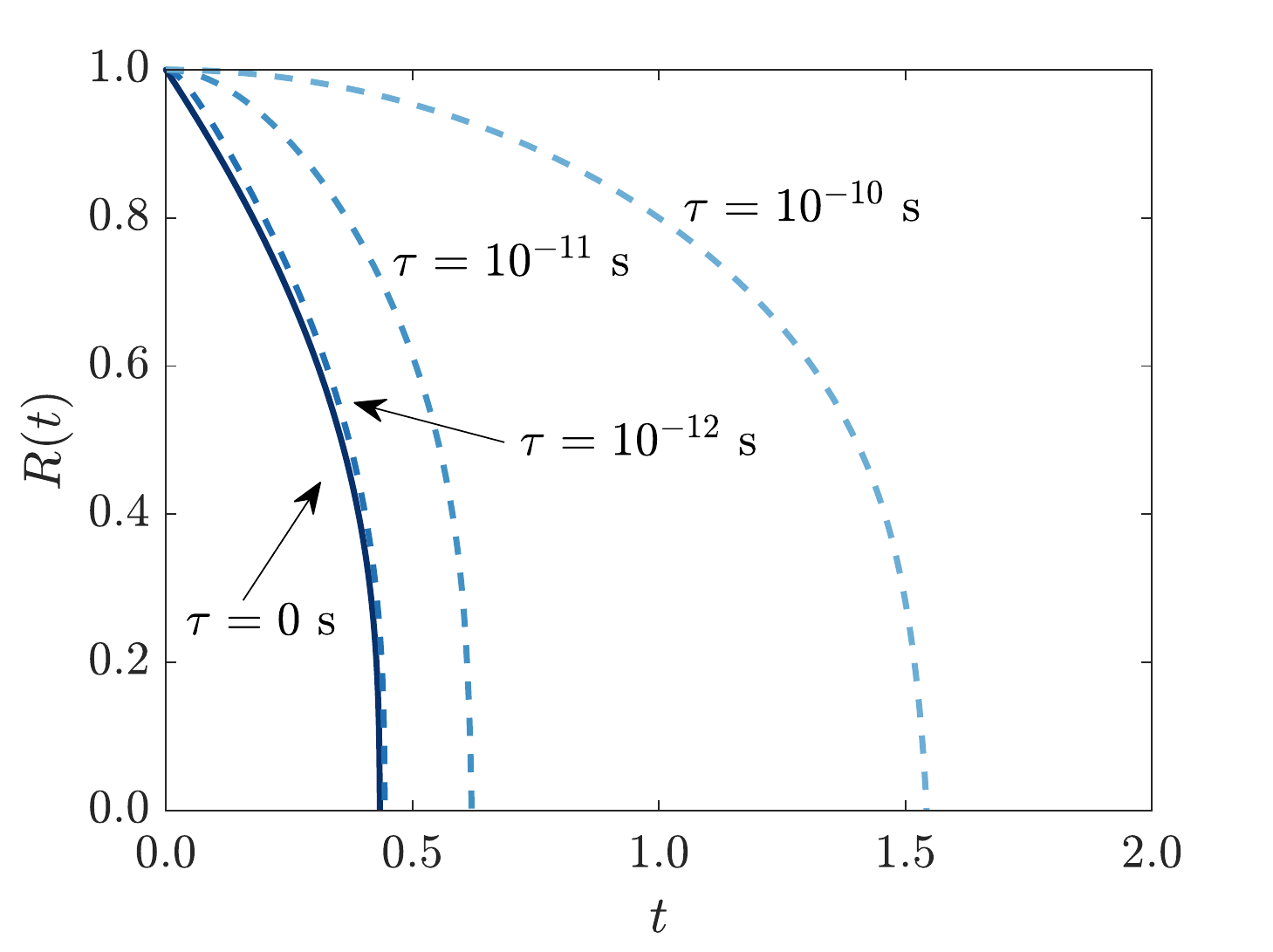}} \\
  \subfigure[Temperature jump]{\includegraphics[width=0.49\textwidth]{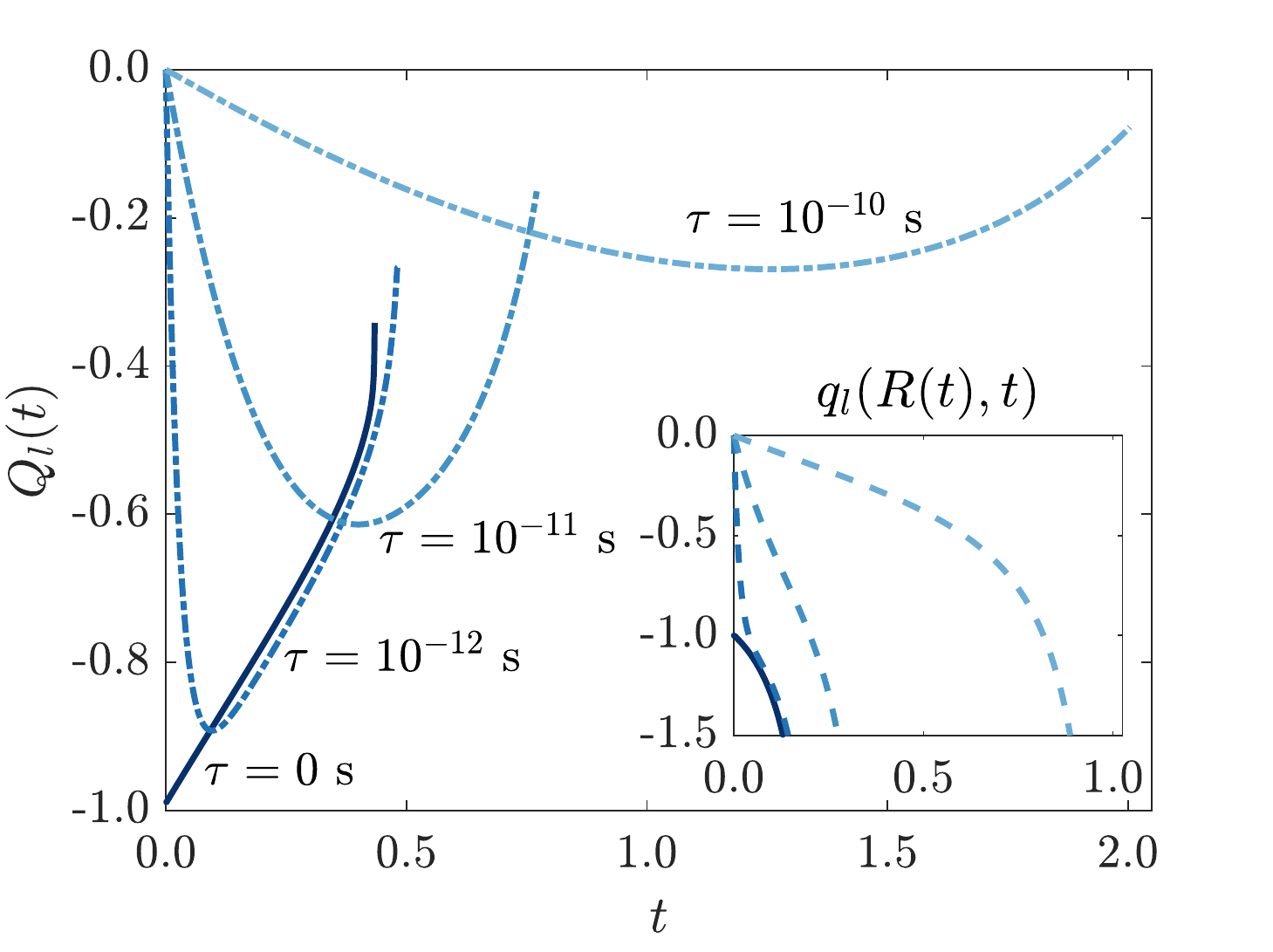}}
  \subfigure[Temperature jump]{\includegraphics[width=0.49\textwidth]{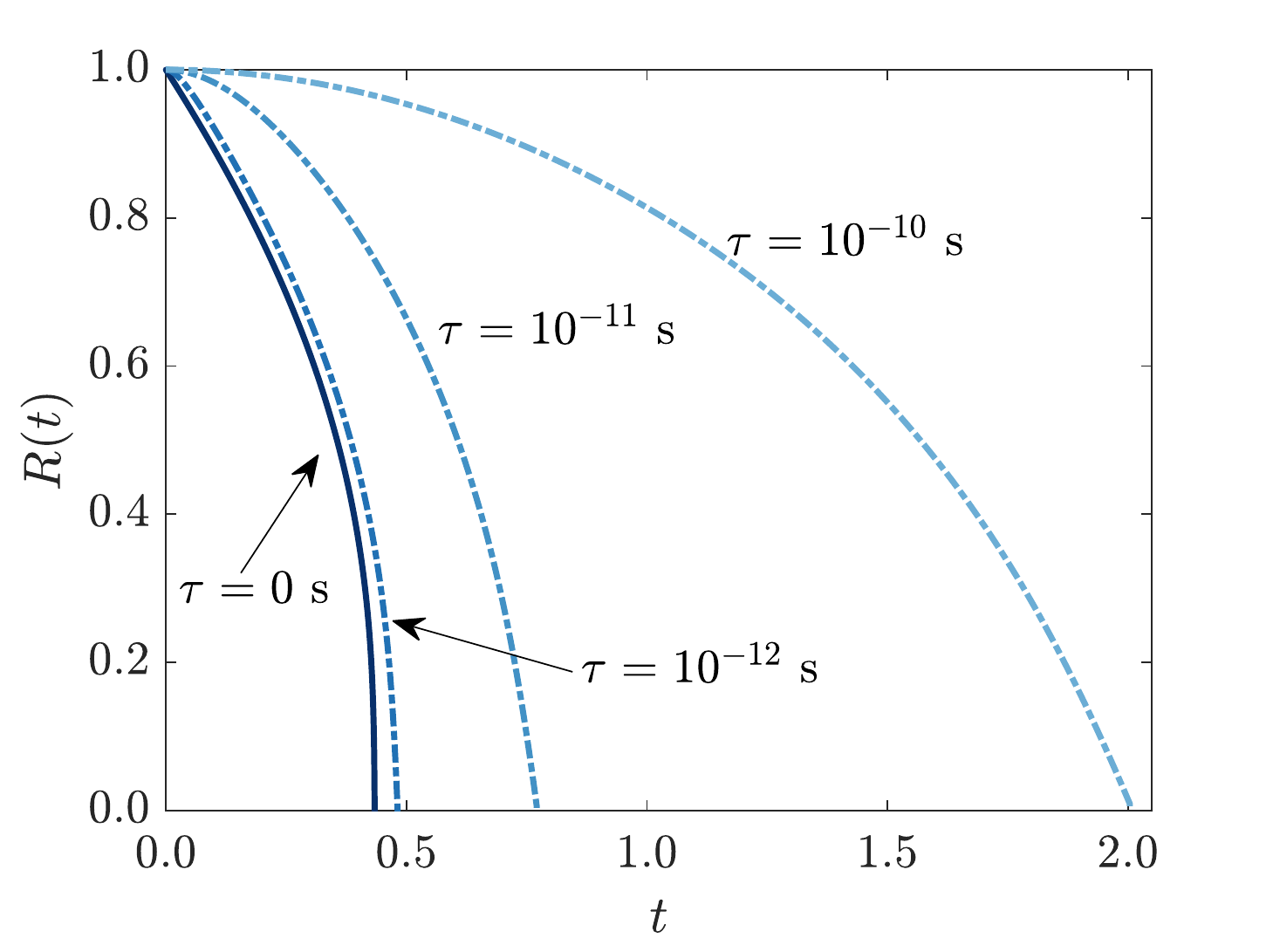}}
  \caption{Influence of the relaxation time $\tau_l = \tau_s = \tau$ on nanoparticle melting in the case of temperature continuity (panels (a) and (b)) and a temperature jump (panels (c) and (d)) across the melt front.  (a) and (c): Evolution of the mean liquid flux (main panel) and liquid flux at the melt front (inset). (b) and (d): Evolution of the position of the melt front. Parameter values are $R_0 = 10$~nm and $\beta = 10$ ($\Delta T \simeq 24$~K).}
  \label{fig:gamma_comp}
\end{figure}

We now explore how the thermal relaxation time affects the melting process for a fixed set of parameter values, assuming that the relaxation times of the solid and the liquid are the same, $\tau_l = \tau_s = \tau$. We take $R_0 = 10$~nm and $\beta = 10$, corresponding to $\Delta T \simeq 24$~K, and examine values of $\tau$ that range from $10^{-12}$~s to $10^{-10}$~s. The effective relaxation parameter $\gamma$ lies between $0.025$ and $2.5$.  The results of numerical simulations carried out with these parameters, along with the case when $\tau = 0$, are shown in Fig.~\ref{fig:gamma_comp}.  Panels (a)--(b) and (c)--(d) show results obtained using the MC-TC and MC-TJ models, respectively.  When $\tau_r = 10^{-12}$~s ($\gamma \simeq 0.025$), the evolution of the mean thermal flux can clearly be divided into the relaxation- and diffusion-dominated time regimes discussed in Sec.~\ref{sec:st_cont}. As the thermal relaxation time is increased, the duration of the first, relaxation-dominated time regime is prolonged. In the case of $\tau_r = 10^{-10}$~s and temperature continuity, the nanoparticle completely melts before the second, diffusion-dominated regime is entered. Although the magnitude of the mean flux for $\tau_r = 10^{-12}$~s and $10^{-11}$~s eventually exceeds the Fourier flux, the inset of Figs.~\ref{fig:gamma_comp} (a) and (c) shows that Fourier's law leads to the largest flux (in magnitude) at the melt front for all times. Thermal relaxation therefore inhibits the transport of thermal energy to the melt front, leading to slower melting kinetics and a melting time that monotonically increases with the relaxation time. 

%As the thermal relaxation time increases from $10^{-12}$~s to $10^{-10}$~s, the melting time increases by roughly a factor of three, suggesting that relaxation plays only a minor role in the melting of tin nanoparticles.  

\begin{figure}
  \centering
  %\subfigure[]{\includegraphics[width=0.49\textwidth]{figs/gamma_comp_Q_100.pdf}}
  \subfigure[Temperature continuity]{\includegraphics[width=0.49\textwidth]{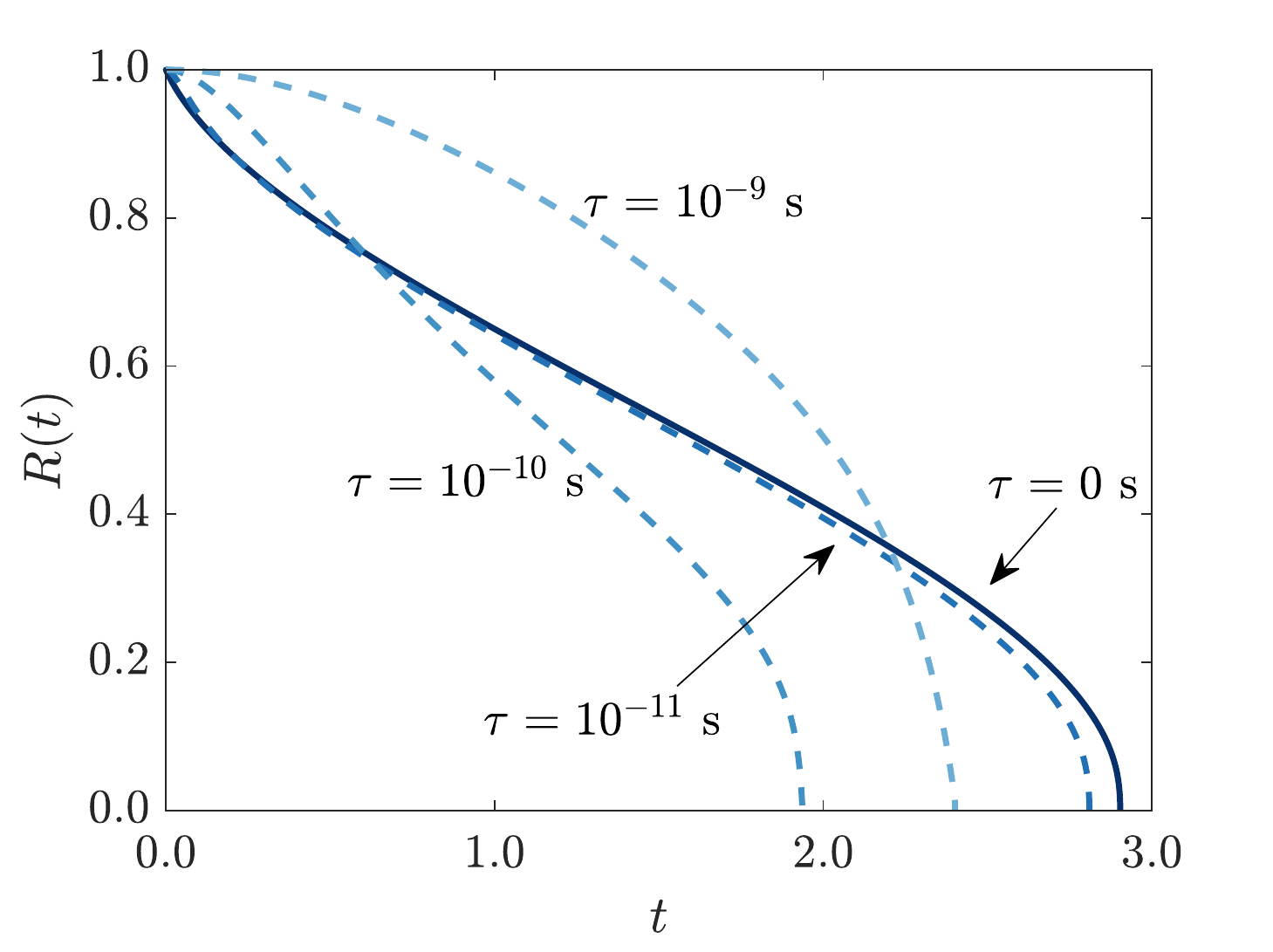}}
  \subfigure[Temperature continuity]{\includegraphics[width=0.49\textwidth]{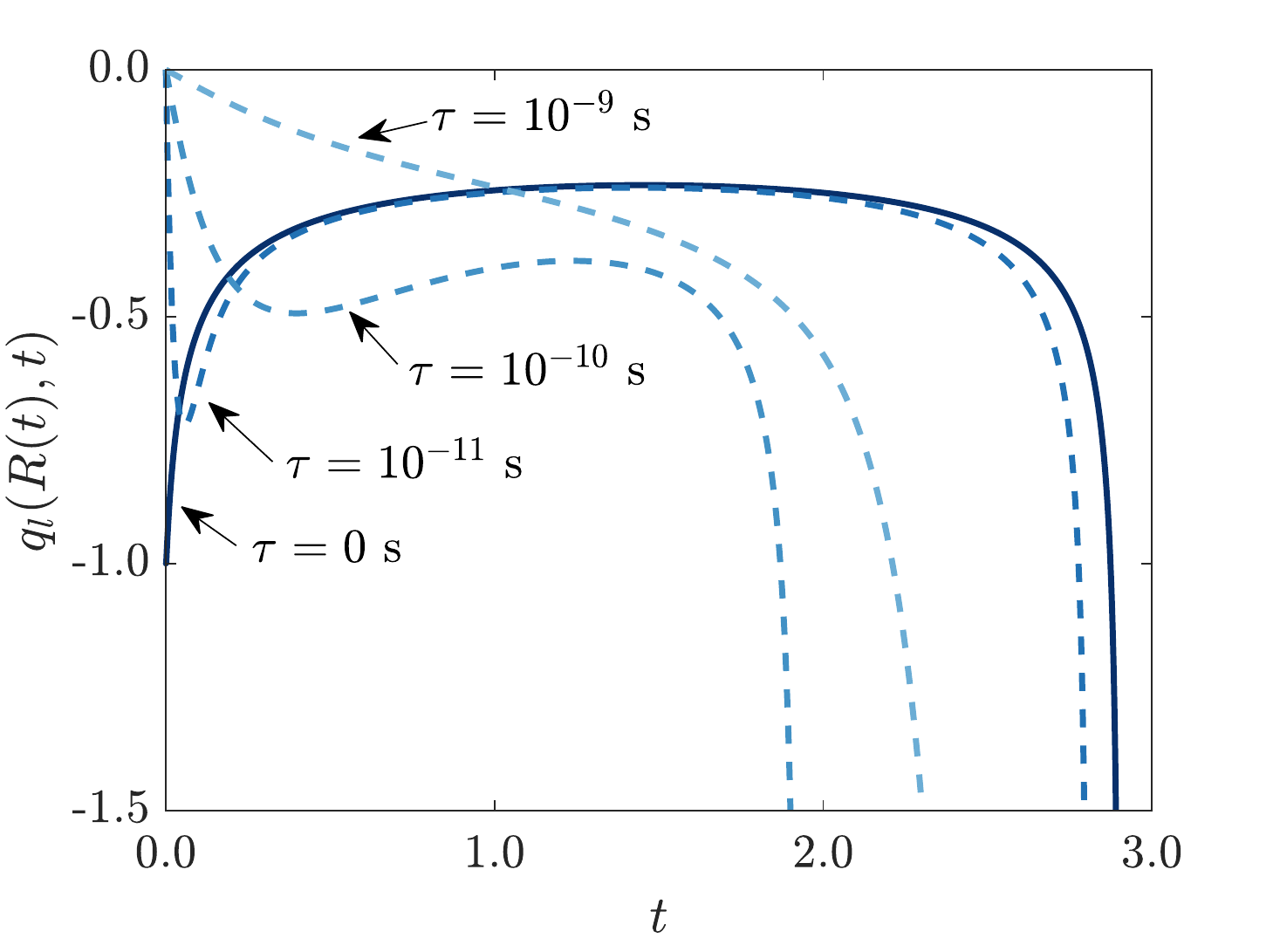}} \\
    \subfigure[Temperature jump]{\includegraphics[width=0.49\textwidth]{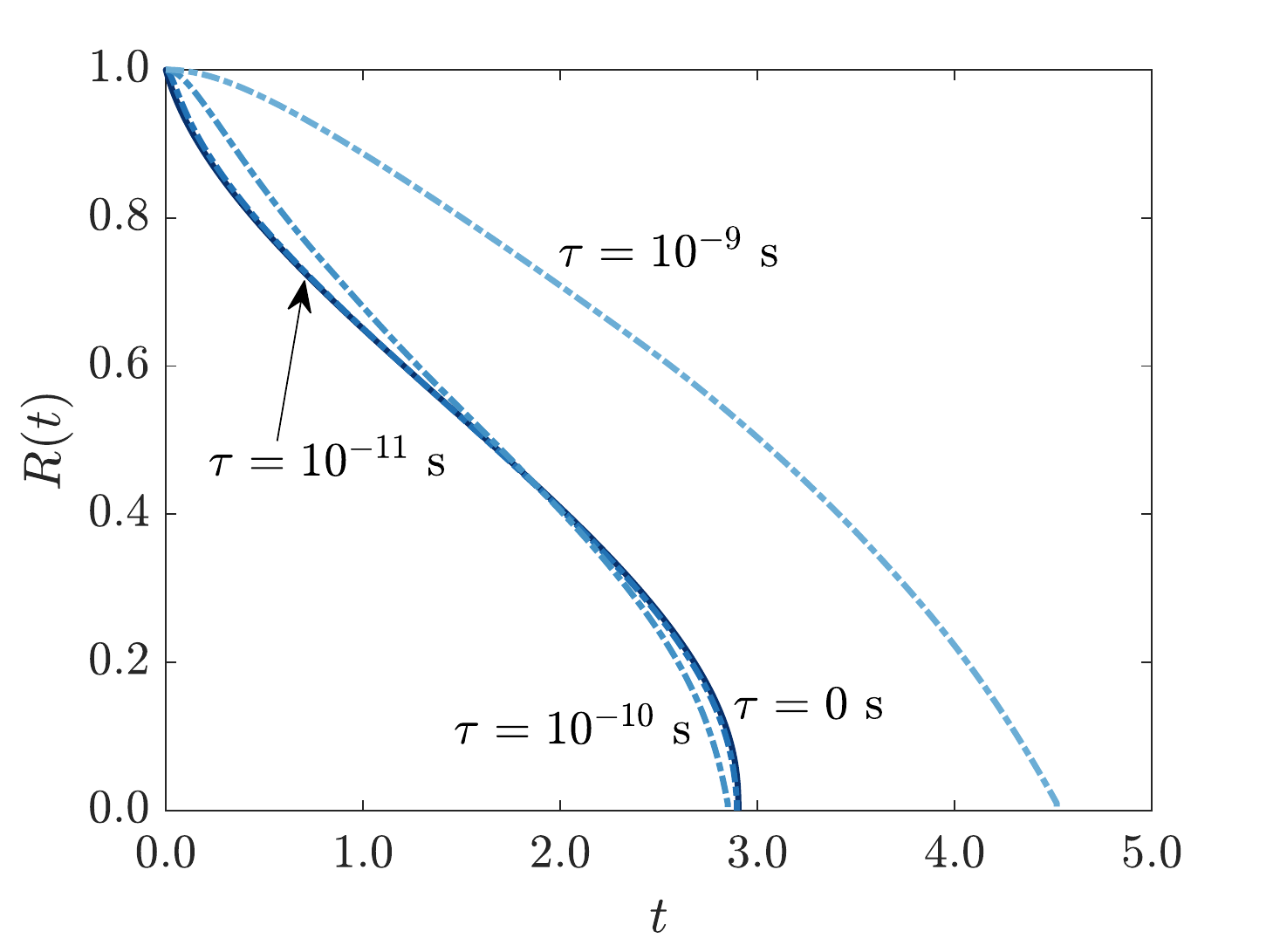}}
  \subfigure[Temperature jump]{\includegraphics[width=0.49\textwidth]{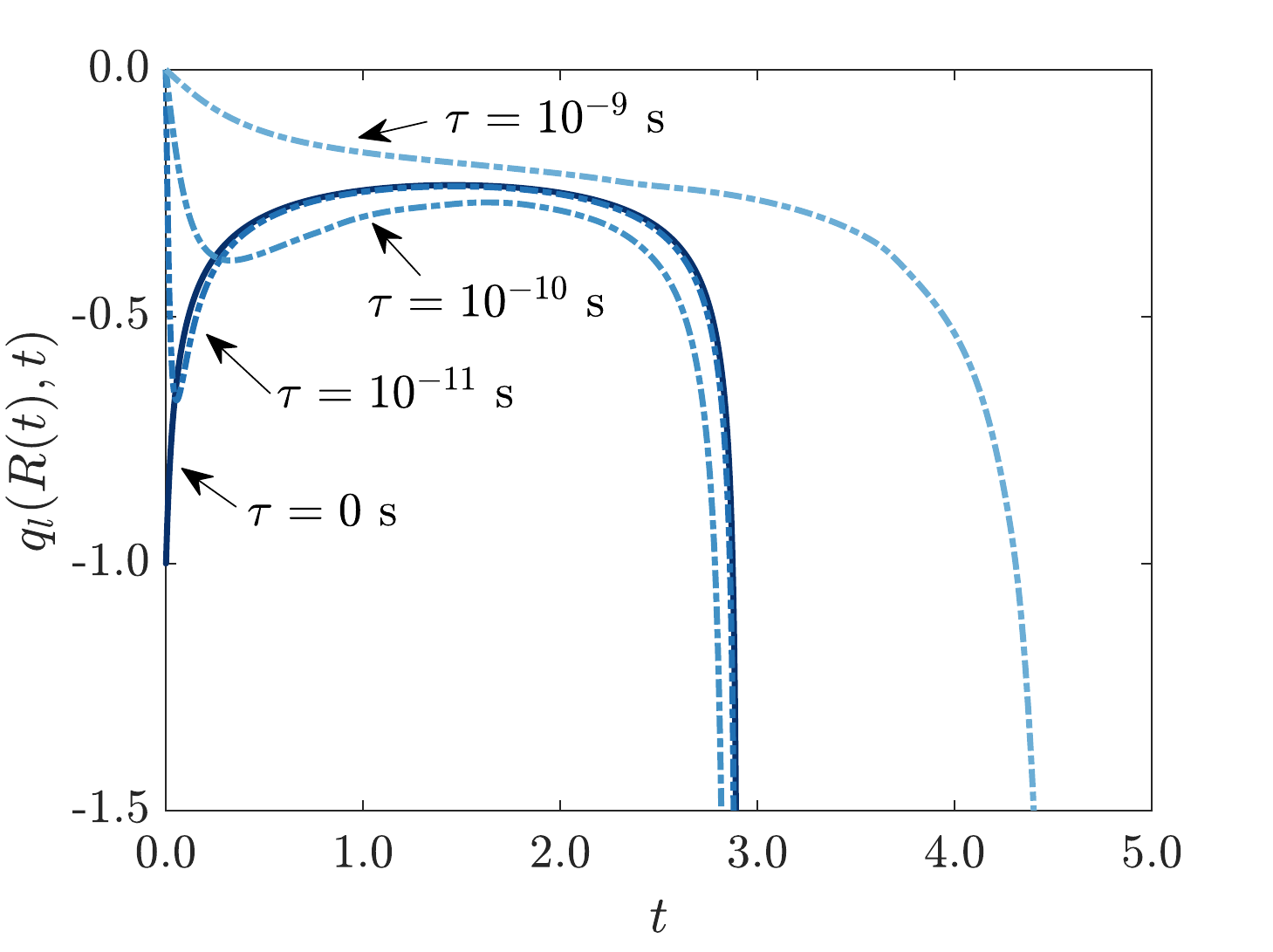}}
  \caption{Influence of the relaxation time $\tau_s = \tau_l = \tau$ on nanoparticle melting in the case of temperature continuity (panels (a) and (b)) and a temperature jump (panels (c) and (d)) across the melt front. Evolution of the position of the melt front ((a) and (c)) and mean flux ((b) and (d)). Parameter values are $R_0 = 100$~nm and $\beta = 10$ ($\Delta T \simeq 24$~K).}
  \label{fig:gamma_comp_100}
\end{figure}

The size of the nanoparticle $R_0$ influences the melting dynamics in a non-trivial manner. This is demonstrated in Fig.~\ref{fig:gamma_comp_100}, which shows the results of numerical simulations carried out using larger nanoparticles, of radius $R_0 = 100$~nm, and relaxation times in the range of $10^{-11}$~s to $10^{-9}$~s. The Stefan number was held at $\beta = 10$. This choice of parameters leads to the same values of the effective relaxation parameter $\gamma$ considered in Fig.~\ref{fig:gamma_comp} for 10~nm nanoparticles, but it increases the Nusselt number by a factor of ten to $\mathcal{N} \simeq 15.7$.  Panels (a)--(b) and (c)--(d) of Fig.~\ref{fig:gamma_comp_100} correspond to the MC-TC and MC-TJ models, respectively.  We find that: (i) thermal relaxation can lead to a faster melting process, (ii) the melting time is a non-monotonic function of the relaxation time, and (iii) the thermal flux at the melt front can be a non-monotonic function of time. These three features are attributed to the competing roles of rapid heat transfer, as characterised by the large Nusselt number, and thermal relaxation. In the case of zero relaxation time, the thermal flux at the melt front (see Fig.~\ref{fig:gamma_comp_100} (b) or (d)) initially decreases in magnitude. This is due to the sudden increase in surface temperature, which, in turn, reduces the driving force behind heat exchange with the environment. In systems with non-zero relaxation time, the sharp increase in surface temperature is dampened by relaxation effects, allowing the magnitude of the flux at the melt front to overshoot and subsequently exceed the Fourier value, leading to a melting process that is faster overall. However, the time at which the overshoot occurs increases with the relaxation time. For sufficiently large values of the relaxation time, no overshoot occurs at all, and the melting time becomes larger compared to the Fourier case; this is shown in Fig.~\ref{fig:gamma_comp_100} (d) when $\tau = 10^{-9}$~s.
%As $\tau_r$ is increased from zero, we find that the melting time decreases until a critical value given by $\tau_r \simeq 2.25\times 10^{-10}$~s is reached, after which the melting time begins to increase with the relaxation time. 
In the case of the 10~nm nanoparticles considered in Fig.~\ref{fig:gamma_comp}, heat is transported away from the surface sufficiently quickly that large increases in temperature and hence decreases in the magnitude of the flux are avoided.

% When the effective relaxation parameter (\emph{i.e.,} the relaxation time) is small and the Nusselt number is large, surface heating occurs on a shorter time scale than thermal relaxation. The rapid increase in temperature near the surface induces a large thermal flux, thereby accelerating the initial melting process. When thermal relaxation becomes relevant, it acts to maintain the large thermal flux induced by efficient heat exchange with the environment. This leads to melting times which decrease with increasing relaxation time. However, as the relaxation time is increased (for fixed Nusselt number), the time scales of thermal relaxation and surface heating become commensurate.  The rapid increase in surface temperature becomes more strongly dampened by relaxaion effects.  Consequently, the thermal flux is reduced, the melting kinetics are slower, and the melting time begins to increase with the relaxation time.

The results shown in Figs.~\ref{fig:gamma_comp} and \ref{fig:gamma_comp_100} indicate that the choice of boundary condition for the Maxwell--Cattaneo model can have a large impact on the solutions. Differences begin to arise in the intermediate stages of the melting process and persist until melting is complete, as clearly shown in Fig.~\ref{fig:all_comp}. The melting kinetics predicted from the MC-TJ model are generally slower than those of the MC-TC model. In the case of the 100~nm nanoparticles shown in Fig.~\ref{fig:gamma_comp_100}, we see that as the relaxation time is decreased from $10^{-9}$~s, the solutions from the MC-TJ model converge to the Fourier model more rapidly compared to the MC-TC model. For instance, the MC-TJ solution obtained with a relaxation time of $10^{-11}$~s is virtually indistinguishable from the Fourier solution, whereas small differences are observed in the MC-TC case. These results are in agreement with those of Greenberg \cite{greenberg1987}, who states that it is the jump condition, not the continuity condition, that leads to solutions that are consistent with those obtained using Fourier's law. Greenberg uses this observation to suggest that the jump condition should be used instead of the continuity condition. However, we do not believe this observation on its own is sufficient to rule out the continuity condition. In fact, the results shown in Fig.~\ref{fig:gamma_comp} for 10~nm nanoparticles appear to contradict Greenberg's statement: MC-TJ solutions for the position of the melt front converge to the Fourier solution slower than the MC-TC solutions. Furthermore, the asymptotic analysis in Sec.~\ref{sec:zero} shows that both models behave in the same manner in the classical limit as $\gamma \to 0$. Rather, we claim that the jump condition is more appropriate to use because it prevents the possibility of supersonic melting, a feature that was not observed in the models of Greenberg.

% \begin{figure}
%   \centering
%   \subfigure[Fourier]{\includegraphics[width=0.49\textwidth]{figs/temp_fourier.png}}
%   \subfigure[Maxwell--Cattaneo]{\includegraphics[width=0.49\textwidth]{figs/temp_mc.png}}
%   \caption{Evolution of the temperature profile in the case of Fourier conduction~(a) and Maxwell--Cattaneo conduction~(b).  To facilitate making a comparison between the models, the evolution is described in terms of the width of the liquid region $1 - R(t)$.  Temperature continuity was assumed when simulating the model with the Maxwell--Cattaneo equation. Parameter values match those of Fig.~\ref{fig:gamma_comp}: $R_0 = 10$~nm and $\beta = 10$.  A value of $\tau_r = 10^{-10}$~s was used in panel (b).}
%   \label{fig:temp}
% \end{figure}

\begin{figure}
  \centering
  \subfigure[Temperature continuity]{\includegraphics[width=0.49\textwidth]{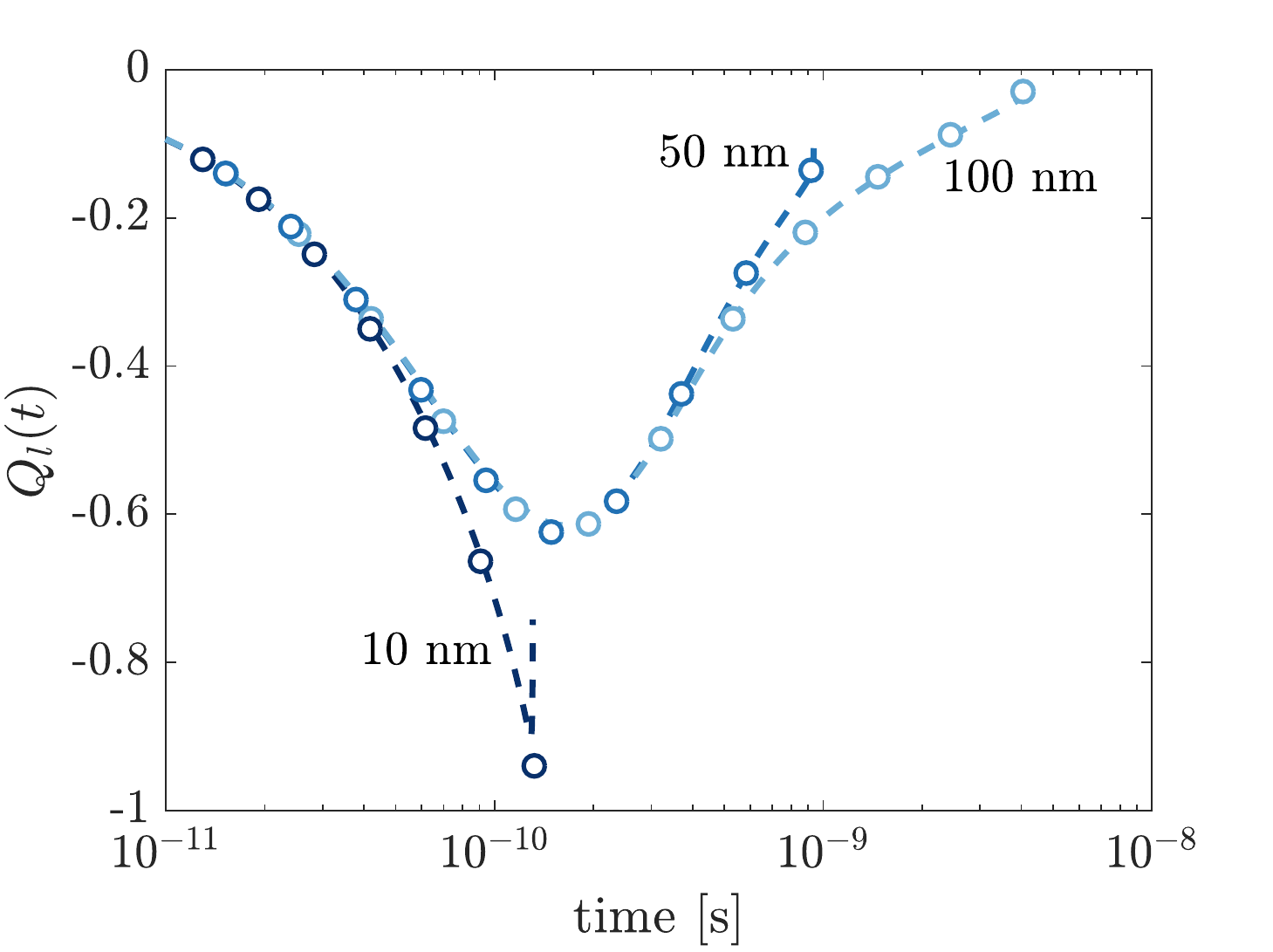}}
  \subfigure[Temperature continuity]{\includegraphics[width=0.49\textwidth]{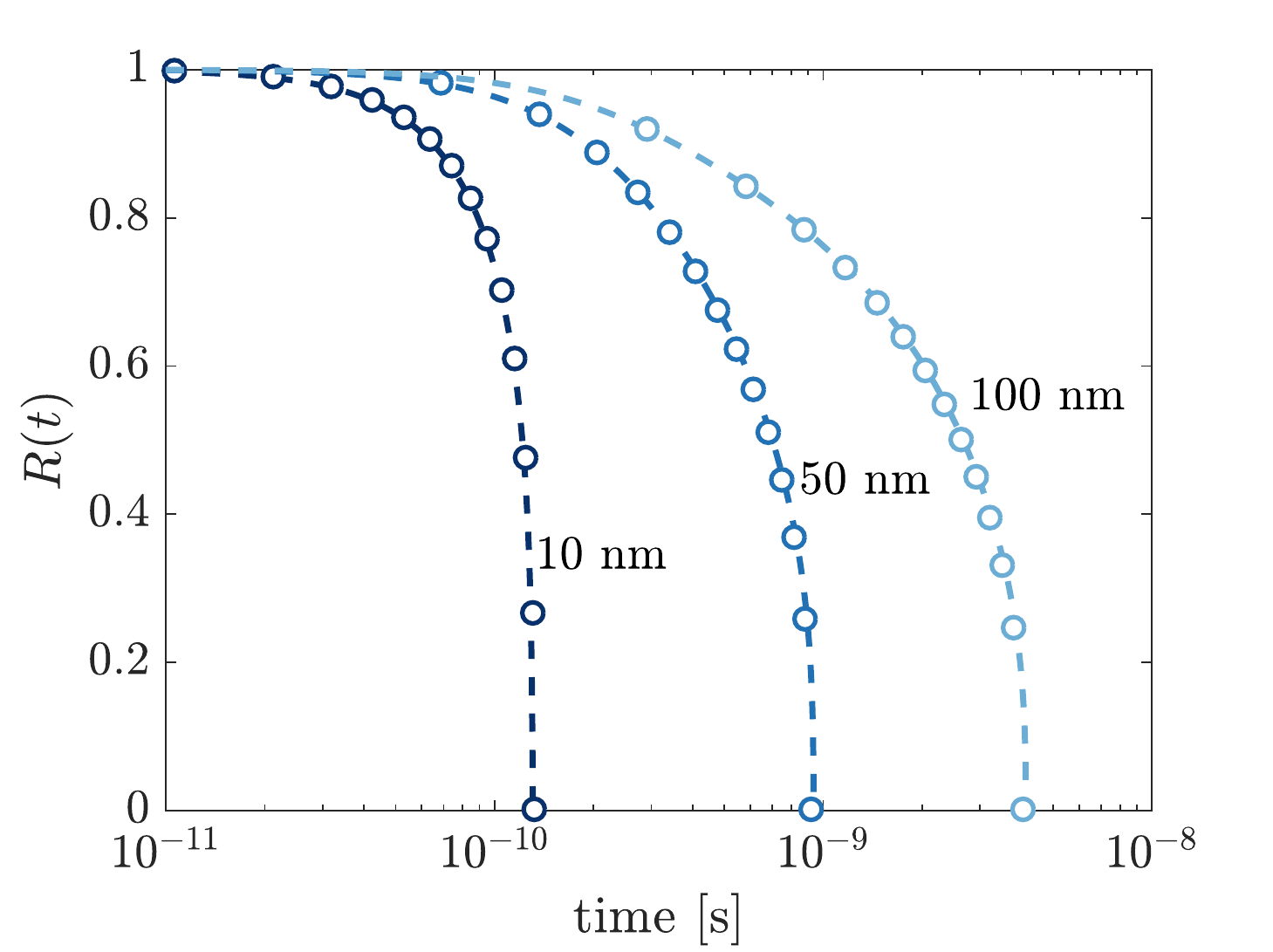}} \\
  \subfigure[Temperature jump]{\includegraphics[width=0.49\textwidth]{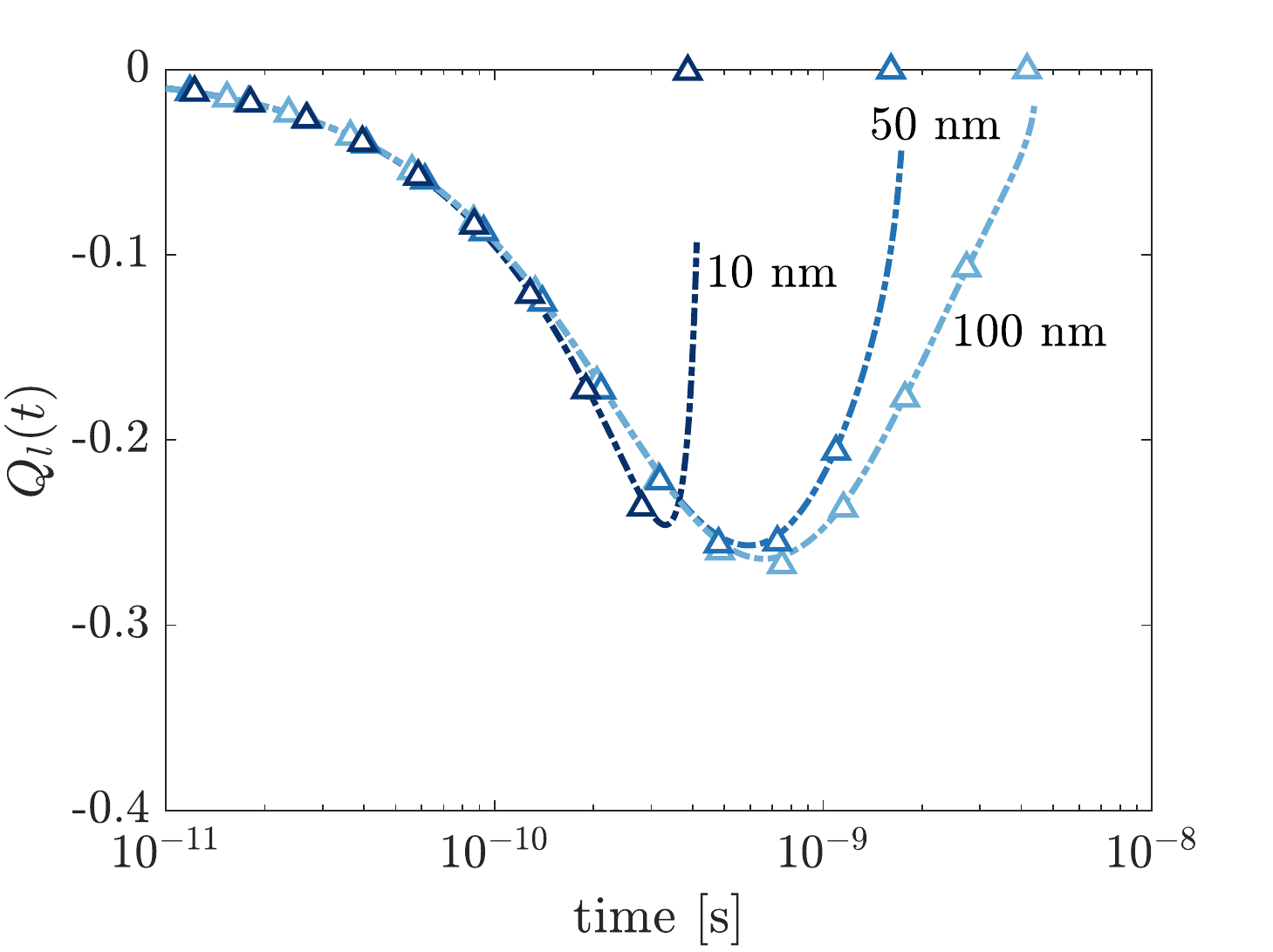}}
  \subfigure[Temperature jump]{\includegraphics[width=0.49\textwidth]{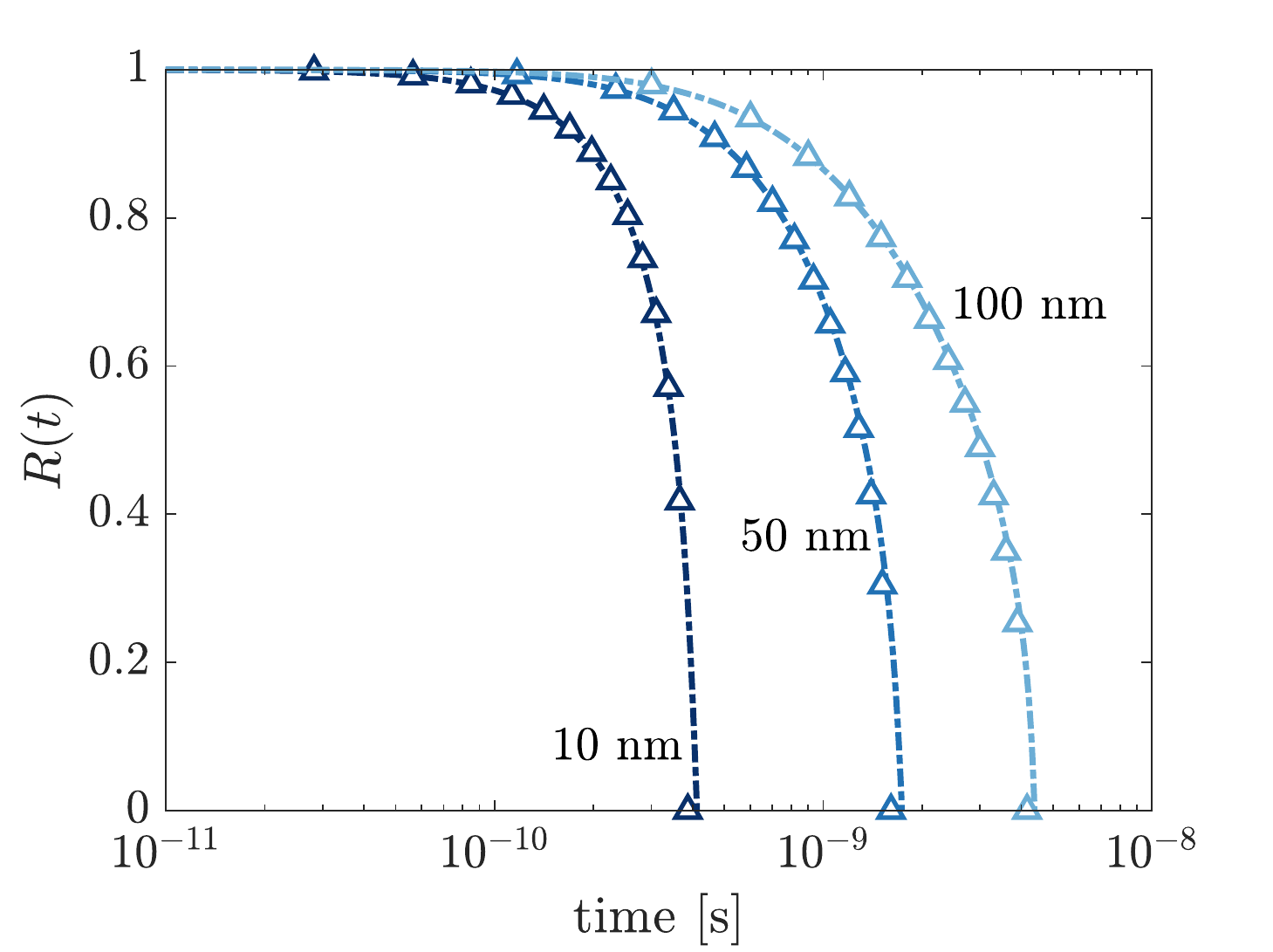}}
  \caption{Comparison of the full model and the asymptotically reduced model for large Stefan numbers and various nanoparticle radii $R_0$. Dashed and dashed-dotted lines denote numerical simulations of the full model \eqref{nd:bulk}--\eqref{nd:newton} with temperature continuity and a temperature jump imposed at the melt front, respectively. Circles and triangles represent numerical simulations of the reduced models given by \eqref{red_s:odes} and \eqref{red:ode_jump}, respectively. 
    (a) Evolution of the mean liquid flux as defined in \eqref{eqn:Q}. 
    (b) Evolution of the position of the solid-liquid interface. 
    Parameter values are $\beta = 50$, corresponding to $\Delta T \simeq 4.7$~K, and $\tau_l = \tau_s = 10^{-10}$~s.}
  \label{fig:large_beta}
\end{figure}

For small temperature differences $\Delta T$ and hence large Stefan numbers, the asymptotically reduced models given by \eqref{red_s:odes} and \eqref{red:ode_jump} can be used to describe nanoparticle melting.  In Fig.~\ref{fig:large_beta}, numerical results obtained from the full model \eqref{nd:bulk}--\eqref{nd:newton} are compared to those of the reduced models.  The Stefan number is set to $\beta = 50$, corresponding to $\Delta T \simeq 4.7$~K, and a relaxation time of $\tau_l = \tau_s = 10^{-10}$~s is used.  Various nanoparticle radii are considered. The position of the melt front and the mean flux are plotted in Fig.~\ref{fig:large_beta} as functions of the dimensional time rather than the dimensionless time because the time scale that is used to non-dimensionalise the model contains a factor of $R_0$ and therefore varies between simulations. From Figs.~\ref{fig:large_beta} (a) and (c), the initial behaviour of the mean liquid flux is seen to be independent of the size of the nanoparticle, which can be verified by re-dimensionalising the small-time solution for the thermal flux \eqref{st:soln_cont_q}, yielding $q_l \sim (h \Delta T)(t / \tau_l)$. The difference in the long-time behaviour of the flux in the MC-TC model is attributed to the varying size of the effective relaxation parameter $\gamma$. We find that $\gamma = 0.50$, $0.10$, and $0.050$ for 10~nm, 50~nm, and 100~nm nanoparticles, respectively. The relatively strong influence of thermal relaxation in 10~nm nanoparticles prevents the second, diffusion-dominated regime of transport from being reached before the solid core is completely transformed into liquid. For all nanoparticle radii, there is excellent agreement between the asymptotically reduced models and the full model.

The role of thermal relaxation in nanoparticles composed of gold and lead can be studied using the parameter values in Table \ref{table:materials}. The maximum value of the heat transfer coefficient for each material can be calculated using the method outlined by Ribera and Myers \cite{ribera2016}. We find that $h_\text{max}$ has the same order of magnitude for each material. For the purpose of this discussion, it is sufficient to set $h_\text{max} = 4.9 \times 10^{9}$~W/(m\unit{2}$\cdot$K) for each material. The melting times of tin, gold, and lead nanoparticles computed using the Fourier, MC-TC, and MC-TJ models and are tabulated in Table \ref{tab:times} for various values of $R_0$ and $\Delta T$. The relaxation time is fixed at $\tau_s = \tau_l = 10^{-10}$~s. Thermal relaxation plays the strongest role in 10~nm nanoparticles that are subject to a 100~K temperature difference, as these conditions lead to the most rapid melting. As the radius of the nanoparticle increases or the temperature difference decreases, both of which lead to a longer melting process, the difference in the Fourier and Maxwell--Cattaneo melting times decreases, implying thermal relaxation plays a weaker role. For each value of $\Delta T$ and $R_0$, the relative difference in melting time is similar across the materials, indicating that thermal relaxation affects each material in roughly the same way. Thus, the results of this section, which are based on parameter values associated with tin, are likely to hold for other materials as well.

\begin{table}
  \centering
  \caption{Dimensional melting times of tin, gold, and lead nanoparticles computed using Fourier's law, $t_\text{F}$, the Maxwell--Cattaneo equation with temperature continuity, $t_\text{MC}^{\text{cont}}$, and the Maxwell--Cattaneo equation with a temperature jump, $t_\text{MC}^{\text{jump}}$, for various radii $R_0$ and temperature differences $\Delta T$. Bold formatting denotes where the largest differences occur.}
  \label{tab:times}
  % \begin{tabular}{|c|ccc|ccc|ccc|}
  %   \hline
  %   & \multicolumn{3}{|c|}{$\Delta T = 1$~K} & \multicolumn{3}{|c|}{$\Delta T = 10$~K} & \multicolumn{3}{|c|}{$\Delta T = 100$~K} \\ \hline \hline
  %   $10$~nm & $t_\text{F}$ [ps]& $t_\text{MC}$ [ps] & $\eta$ & $t_\text{F}$ [ps] & $t_\text{MC}$ [ps] & $\eta$ &  $t_\text{F}$ [ps] & $t_\text{MC}$ [ps] & $\eta$ \\ \hline
  %   Tin  & 110 & 248  & 1.26 & 30.4 & 88.4  & 1.91 & 5.50  & 30.1  & 4.46 \\
  %   Gold & 99.6 & 272  & 1.73 & 35.2 & 121  & 2.43 & 7.17  & 41.6  & 4.81 \\
  %   Lead & 54.7 & 166  & 2.03 & 19.1 & 67.6 & 2.54 & 4.43  & 26.7  & 5.04 \\
  %   \hline \hline
  %       $100$~nm & $t_\text{F}$ [ns]& $t_\text{MC}$ [ns] & $\eta$ & $t_\text{F}$ [ns] & $t_\text{MC}$ [ns] & $\eta$ &  $t_\text{F}$ [ns] & $t_\text{MC}$ [ns] & $\eta$ \\ \hline
  %   Tin  & 12.2 & 12.0 & 0.011 & 2.25 & 1.94 & 0.139 & 0.329 & 0.230 & 0.300 \\
  %   Gold & 6.73 & 6.86 & 0.018 & 1.71 & 1.65 & 0.038 & 0.273 & 0.242 & 0.113 \\
  %   Lead & 9.75 & 9.61 & 0.015 & 2.21 & 1.87 & 0.155 & 0.372 & 0.279 & 0.250 \\
  %   \hline
  % \end{tabular}
  \begin{tabular}{|c|ccc|ccc|ccc|}
    \hline
    & \multicolumn{3}{|c|}{$\Delta T = 1$~K} & \multicolumn{3}{|c|}{$\Delta T = 10$~K} & \multicolumn{3}{|c|}{$\Delta T = 100$~K} \\ \hline \hline
    $10$~nm & $t_\text{F}$ [ps]& $t_\text{MC}^{\text{cont}}$ [ps] & $t_\text{MC}^{\text{jump}}$ [ps] & $t_\text{F}$ [ps] & $t_\text{MC}^{\text{cont}}$  [ps] & $t_\text{MC}^{\text{jump}}$ [ps] &  $t_\text{F}$ [ps] & $t_\text{MC}^{\text{cont}}$  [ps] & $t_\text{MC}^{\text{jump}}$ [ps] \\ \hline
    Tin  & 110  & 245 & 267 & 29.7 & 87.4 & 107  & 5.30  & 29.9  & 46.2 \\
    Gold & 106  & 264 & 275 & 34.3 & 117  & 128  & 6.78  & 40.6  & 51.1\\
    Lead & 53.5 & 161 & 187 & 17.8 & 65.8 & 89.8 & \textbf{3.97}  & \textbf{26.5}  & \textbf{47.0}\\
    \hline \hline
    $100$~nm & $t_\text{F}$ [ns]& $t_\text{MC}^{\text{cont}}$ [ns] & $t_\text{MC}^{\text{jump}}$ [ns] & $t_\text{F}$ [ns] & $t_\text{MC}^{\text{cont}}$  [ns] & $t_\text{MC}^{\text{jump}}$ [ns] &  $t_\text{F}$ [ns] & $t_\text{MC}^{\text{cont}}$  [ns] & $t_\text{MC}^{\text{jump}}$ [ns] \\ \hline
    Tin  & 12.1 & 12.0 & 12.2 & 2.24 & 1.92 & 2.22 & 0.327 & 0.230 & 0.413 \\
    Gold & 6.76 & 6.81 & 6.97 & 1.70 & 1.63 & 1.80 & 0.270 & 0.241 & 0.366 \\
    Lead & 9.66 & 9.51 & 9.80 & 2.18 & 1.84 & 2.19 & 0.364 & 0.278 & 0.473 \\
    \hline
  \end{tabular}

\end{table}

Finally, we address one of the major criticisms of the Maxwell--Cattaneo equation, which is that it can lead to negative entropy production, in violation of the second law of thermodynamics.  The dimensionless entropy generation $\dot{S}$ can be written as \cite{bright2009}
\begin{align}
  \dot{S} = -\frac{q}{\left[\theta(1 - \ell) + T\right]^2}\pd{T}{r}.
\end{align}
Thus, positive entropy generation requires thermal energy to travel in the direction of negative temperature gradient. When the liquid temperature has a maximum near the melt front, as in Fig.~\ref{fig:temp} (c), there are regions of entropy production, indicating a breakdown of the model.
%However, this breakdown occurs late in the melting process, when the radius of the solid core has decreased to roughly 20\% of its initial value. In dimensional terms, this corresponds to a radius of 2~nm. On such length scales, the validity of the continuum approximation also comes into question. Up to this point, the Maxwell--Cattaneo equation provides a thermodynamically consistent description of nanoparticle melting.
Determining explicit conditions that guarantee positive entropy production is difficult. In the case of a large Stefan number, it is straightforward to show that the asymptotically reduced model in the case of temperature continuity, given by \eqref{red_s:odes}, will always produce entropy, therefore ensuring the second law of thermodynamics is satisfied during the melting process. Future studies of nanoparticle melting can explore using alternative models of non-Fourier heat conduction to avoid the issue of negative entropy production.

%-----------------------------------------------------

\section{Conclusions}\label{sec:conclusion}

The melting of spherical nanoparticles is studied using a non-classical model of heat conduction, based on the Maxwell--Cattaneo equation, that accounts for the finite thermal relaxation time of materials. By seeking asymptotic and numerical solutions to the model using realistic parameter values, we show that the jump boundary condition for the temperature produces the most physically meaningful results. In particular, the jump condition avoids the possibility of supersonic melting, whereby the speed of the melt front exceeds the speed of heat propagation. However, deriving the correct jump condition is not straightforward and requires the formulation of a diffuse interface model when capturing variations in the thermal conductivity. Similar considerations may be required for other non-Fourier conduction laws as well.
%produces non-physical results. In particular, the jump condition leads to an effective nanoparticle melting temperature that is much greater than the bulk melting temperature, which contradicts the findings of several experimental studies. Furthermore, it is not generally possible to recover the Fourier-based solution from the Maxwell--Cattaneo model when the jump condition is imposed. Our results highlight important inadequacies in the jump condition and demonstrate that it is not appropriate to use in models of phase change that are based on the Maxwell--Cattaneo equation.

We find there is a non-trivial interplay between heat exchange with the environment and thermal relaxation. Depending on the size of the nanoparticle, thermal relaxation can either decelerate or accelerate the overall melting process relative to the case of Fourier conduction. The relevance of thermal relaxation generally increases as the nanoparticle radius decreases or the imposed temperature difference increases, both of which lead to melting processes that occur more rapidly. For the parameter range and materials considered here, thermal relaxation is found to modify the melting time by more than a factor of 10. We can conclude that thermal relaxation is an important process to capture in models of nanoparticle melting and will likely influence the dynamics of phase change in other technologically relevant systems as well.

We have previously studied the solidification kinetics that arise when heat conduction is described by the Guyer--Krumhansl (GK) equation, which is an extension to the Maxwell--Cattaneo equation that considers the mean free path of phonons \cite{hennessy2018}. In the limit of large relaxation time, we found that the rate of solidification is reduced compared to the Fourier prediction, which is similar to the difference in melting dynamics observed in this paper when Fourier's law is replaced by the Maxwell--Cattaneo equation. However, the physical mechanisms responsible for the decrease in phase-change kinetics are not the same in the two models. In the case of the GK equation, non-classical transport mechanisms act in conjunction to recover a form of Fourier's law with a very small effective thermal conductivity. In this paper, we find that the decreased rate of melting is mainly driven by thermal relaxation affecting the influx of thermal energy from the surrounding environment and the temperature jump at the interface, neither of which were considered in the analysis of the GK equation. These differences illustrate the need for a careful choice of model when considering phase change on small length and time scales.

Given the range of mathematical descriptions for nanoparticle melting, an important area of future research should focus on model validation.  Ideally, this would involve a comparison with experimental data for the temperature profiles and motion of the solid-liquid interface. For instance, comparing the small-time melting kinetics observed experimentally to those predicted theoretically could provide useful insights into the most suitable condition to impose at the outer boundary of the nanoparticle. However, due to the small length and time scales involved in nanoparticle melting, such data is likely to be difficult to acquire with current technology.  Thus, it may be insightful to use molecular dynamics simulations to generate data that can assist with model validation.

%A key shortcoming of the proposed model is that it can violate the second law of thermodynamics and lead to negative production of entropy.
% Numerical simulations show that negative entropy production occurs when the speed of the melt front exceeds the finite speed of heat propagation. Such behaviour is only expected to take place near the end of the melting process when the radius of the solid core becomes a few nanometers in size. Until this point, the Maxwell--Cattaneo equation is able to provide a thermodynamically consistent description of nanoscale heat transport.

%-----------------------------------------------------

\section*{Acknowledgements}
The authors thank Brian Wetton and Sonia Fernandez-Mendez for helpful discussions regarding the numerical solution of the governing equations, as well as the anonymous reviewers for insightful comments about the model. This project has received funding from the European Union's Horizon 2020 research and innovation programme under the Marie Sk{\l}odowska-Curie grant agreement No. 707658. MCS acknowledges that the research leading to
these results has received funding from `la Caixa' Foundation. TGM acknowledges financial support from the Ministerio de Ciencia e Innovaci\'on grant MTM2017-82317-P. The authors have been partially funded by the CERCA Programme of the Generalitat de Catalunya.

%-----------------------------------------------------

\begin{appendix}
  
  \section{Derivation of jump and interfacial conditions from a diffuse-interface model}
  \label{app:diffuse}

  The temperature jump that arises when using the Maxwell--Cattaneo law to model heat conduction is the result of assuming the solid-liquid interface is infinitely thin \cite{sobolev1996}. In reality, however, the transition from the solid to liquid phase occurs over a region with finite width. The diffuse nature of the solid-liquid interface leads to profiles for the temperature, flux, and material properties that vary smoothly from one phase to the other. The jump conditions for a sharp-interface model can be derived from a diffuse-interface model in the limit where the interfacial width asymptotically tends to zero. By deriving the jump conditions in this manner, phase-dependent material properties can be considered in the model. The purpose of this appendix is to propose a simple diffuse-interface model of nanoparticle melting based on the Maxwell--Cattaneo equation. A sharp-interface limit will then be taken in order to extend the jump conditions that were derived by applying the Rankine--Hugoniot condition to a sharp-interface model with constant material properties. 

  % The diffuse-interface model that we consider is based on that of Wang \etal\cite{wang1993}.

  We let $\phi$ denote an order parameter which is conceptually equivalent to the liquid volume fraction. Therefore, regions of pure solid and liquid are characterised by $\phi = 0$ and $\phi = 1$, respectively. The radial positions $R(t)$ where $\phi(R(t),t) = 1/2$ are assumed to correspond to the position of a hypothetical sharp interface that separates the solid and liquid regions. The internal energy per unit mass of the system is written as in Greenberg \cite{greenberg1987}:
  \begin{align}
    u(T,\phi) = c(T - T_m^*) + L_m \phi.
  \end{align}
  The bulk equations for the temperature $T$ and thermal flux $q$ are given by
  \subeq{
    \begin{align}
      \rho c \pd{T}{t} + \rho L_m \pd{\phi}{t} + \frac{1}{r^2}\pd{}{r}\left(r^2 q\right) &= 0, \\
      \tau(\phi) \pd{q}{t} + q + k(\phi)\pd{T}{r} &= 0,
    \end{align}
  }
  $\tau(\phi)$ and $k(\phi)$ are the phase-dependent thermal relaxation time and conductivity, respectively, that satisfy $\tau(0) = \tau_s, \tau(1) = \tau_l$, $k(0) = k_s$ and $k(1) = k_l$. For our purposes, it is sufficient to assume that the order parameter is given by
  \begin{align}
    \phi(r,t) = \frac{1}{2}\left[\tanh\left(\frac{r - R(t)}{\epsilon}\right) + 1\right],
  \end{align}
  where $\epsilon$ is the width of the diffuse interfacial layer, although this choice is ultimately immaterial.
  % Furthermore, $p$ is a monotonically increasing interpolation function satisfying $p(0) = 0$, $p(1/2) = 1/2$, and $p(1) = 1$; $g$ is a double-well potential given by $g = \phi^2(1 - \phi)^2$. A specific form of $p$ is not required here. 
  By non-dimensionalising the bulk equations as in Sec.~\ref{sec:nondim}, we obtain
  \subeq{
  \label{app:nd_bulk}
  \begin{align}
   \beta^{-1}\pd{T}{t} + \pd{\phi}{t} + \frac{1}{r^2}\pd{}{r}\left(r^2 q\right) &= 0, \label{app:nd_bulk_T}\\
    \gamma \tau(\phi) \pd{q}{t} + q + \mathcal{N}^{-1} k(\phi)\pd{T}{r} &= 0, \label{app:nd_bulk_q}
    %\frac{\delta^2}{r^2}\pd{}{r}\left(r^2\pd{\phi}{r}\right) + \frac{2 \delta}{3}\left(\frac{T}{\theta \ell} - 1\right) p'(\phi) - 2 g'(\phi) &= 0, \label{app:nd_bulk_phi}
   \end{align}
   }
   where now $\tau(0) = \tau_r$, $\tau(1) = 1$, $k(0) = k_r$, and $k(1) = 1$.  This non-dimensionalisation introduces the parameter $\delta = \epsilon / R_0$, corresponding to the non-dimensional width of the diffuse solid-liquid interface. Molecular dynamics simulations predict that the width of the solid-liquid interface for copper and aluminium is on the order of a few Angstroms \cite{frolov2009, jesson2000}. Thus, $\delta$ is expected to be small for nanoparticles of radius 10~nm or larger. 

A sharp-interface model with compatible jump conditions can be derived from the asymptotic limit as $\delta \to 0$. Physically, this limit collapses the diffuse interfacial layer, which has a dimensionless width that is $O(\delta)$ in size, to a sharp interface located at $r = R(t)$. Naively taking $\delta \to 0$ shows that $\phi = 0$ for $r < R(t)$ and $\phi = 1$ for $r > R(t)$. This leads to bulk equations for the temperature and flux given by
\subeq{
  \label{app:nd_si}
\begin{align}
  \beta^{-1} \pd{T_s}{t} + \frac{1}{r^2}\pd{}{r}\left(r^2 q_s \right) &= 0, \qquad 0 < r < R(t), \label{app:nd_ce_s} \\
  \gamma \tau_r \pd{q_s}{t} + q_s + k_r \mathcal{N}^{-1} \pd{T_s}{r} &= 0, \qquad 0 < r < R(t), \label{app:nd_c_s} \\
  \beta^{-1} \pd{T_l}{t} + \frac{1}{r^2}\pd{}{r}\left(r^2 q_l \right) &= 0, \qquad R(t) < r < 1, \label{app:nd_ce_l} \\
  \gamma \pd{q_l}{t} + q_l + \mathcal{N}^{-1} \pd{T_l}{r} &= 0, \qquad R(t) < r < 1, \label{app:nd_mc_l}
\end{align}
}
which are identical to \eqref{nd:bulk}. To derive a consistent set of boundary conditions for \eqref{app:nd_si} at the sharp interface $r = R(t)$, we must resolve the solution in the diffuse interfacial layer. We thus introduce the change of variable $r = R(t) + \delta \xi$ and $t = t'$ in the bulk equations \eqref{app:nd_bulk} to obtain (upon dropping the prime)
  \subeq{
  \label{app:nd_bulk2}
  \begin{align}
    \beta^{-1}\left(\pd{T}{t} - \delta^{-1}\dot{R}\pd{T}{\xi}\right) + \pd{\phi}{t} - \delta^{-1} \dot{T}\pd{\phi}{\xi} + \frac{\delta^{-1}}{(R + \delta \xi)^2}\pd{}{\xi}\left[(R + \delta \xi)^2 q\right] &= 0, \label{app:nd_bulk_T2}\\
    \gamma \tau(\phi) \left(\pd{q}{t} - \delta^{-1} \dot{R}\pd{q}{\xi}\right) + q + \delta^{-1} \mathcal{N}^{-1} k(\phi)\pd{T}{\xi} &= 0. \label{app:nd_bulk_q2}
    %\frac{1}{(R + \delta \xi)^2}\pd{}{\xi}\left[(R + \delta \xi)^2\pd{\phi}{\xi}\right] + \frac{2\delta}{3}\left(\frac{T}{\theta \ell} - 1\right) p'(\phi) - 2 g'(\phi) &= 0. \label{app:nd_bulk_phi2}
   \end{align}
 }
 The solution is now expanded as an asymptotic series of the form $T = T^{(0)} + O(\delta)$ and $q = q^{(0)} + O(\delta)$. The matching conditions are given by
 \subeq{
 \begin{alignat}{2}
   T^{(0)}(\xi \to -\infty, t) &= T_s(R(t),t), &\quad T^{(0)}(\xi \to \infty,t) &= T_l(R(t),t), \label{app:match_T}\\
   q^{(0)}(\xi \to -\infty, t) &= q_s(R(t),t), &\quad q^{(0)}(\xi \to \infty,t) &= q_l(R(t),t). \label{app:match_q}
 \end{alignat}
}
The leading-order contributions to the bulk equations for the temperature and flux, \eqref{app:nd_bulk_T2} and \eqref{app:nd_bulk_q2}, are
  \subeq{
  \begin{align}
   -\beta^{-1} \dot{R}\pd{T^{(0)}}{\xi} -\dot{R} \pd{\phi}{\xi} + \pd{q^{(0)}}{\xi} &= 0, \label{app:si_T} \\
    -\gamma \tau(\phi) \dot{R} \pd{q^{(0)}}{\xi} + \mathcal{N}^{-1} k(\phi)\pd{T^{(0)}}{\xi} &= 0.
   \label{app:si_q}
   \end{align}
   }
By solving \eqref{app:si_q} for $\pdf{q^{(0)}}{\xi}$ and substituting the result into \eqref{app:si_T}, we find that the temperature across the diffuse interface varies according to
\begin{align}
  \pd{T^{(0)}}{\xi} = \beta \left(\frac{\beta k(\phi)}{\mathcal{N} \gamma \tau(\phi) \dot{R}^2}-1\right)^{-1}\pd{\phi}{\xi}.
  \label{app:T_xi}
\end{align}
The temperature jump across the solid-liquid interface can be obtained by integrating \eqref{app:T_xi} across the domain and using the matching conditions in \eqref{app:match_T} to obtain
\begin{align}
  T_l(R,t) - T_s(R,t) =
  % \beta \int_{-\infty}^{+\infty}\left(\frac{\beta k(\phi^{(0)})}{\mathcal{N} \beta \tau(\phi^{(0)}) \dot{R}^2}-1\right)^{-1}p'(\phi^{(0)})\pd{\phi^{(0)}}{\xi}\,\d \xi =
  \beta \int_{0}^{1}\left(\frac{\beta k(\Phi)}{\mathcal{N} \gamma \tau(\Phi) \dot{R}^2}-1\right)^{-1}\,\d \Phi.
  \label{app:jump}
\end{align}
The temperature profile in the interfacial layer can be written as
\begin{align}
  T^{(0)}(\xi, t) = T_s(R(t),t) + \beta \int_{0}^{\phi(\xi,t)} \left(\frac{\beta k(\Phi)}{\mathcal{N} \gamma \tau(\Phi) \dot{R}^2}-1\right)^{-1}\,\d \Phi.
  \label{app:T_sol}
\end{align}
The temperature of the solid at the melt front, $T_s(R(t),t)$, can be determined by imposing an admissibility condition or through a solvability condition that is derived from prescribing a governing equation for the order parameter \cite{greenberg1987, caginalp1989}. Here, it will be assumed that the temperature is equal to the Gibbs--Thomson melting temperature when $\phi = 1/2$, \emph{i.e.}, at the position of the sharp interface. By combining the jump condition \eqref{app:jump} with $T^{(0)}(0,t) = T_m(R(t))$, the temperature of the solid and liquid at the melt front, $T_s(R(t),t)$ and $T_l(R(t),t)$, can be obtained.

Taking $\gamma \to 0$ in \eqref{app:T_sol} shows that the temperature is uniform across the interfacial layer in the classical limit. The matching conditions \eqref{app:match_T} imply that $T_s(R,t) = T_l(R,t) = T^{(0)}(t) = T_m(R)$, corresponding to the case of temperature continuity across the melt front,

In the case of constant thermal conductivity and relaxation time, the jump condition \eqref{app:jump} is identical to \eqref{eqn:const_T_jump}, the latter of which was derived using the Rankine--Hugoniot condition. If the thermal conductivity and relaxation time are piecewise constant, then the interfacial conditions for the solid and liquid temperatures are given by
\subeq{
  \label{app:T_inter}
  \begin{align}
    T_s(R(t),t) &= \theta \ell \left(1 - \frac{1}{R}\right) - \frac{\beta}{2}\left(\frac{\beta k_r}{\mathcal{N} \gamma \tau_r \dot{R}^2} - 1 \right)^{-1}\\
    T_l(R(t),t) &= \theta \ell \left(1 - \frac{1}{R}\right) + \frac{\beta}{2}\left(\frac{\beta}{\mathcal{N} \gamma \dot{R}^2}-1\right)^{-1}.
  \end{align}
}
Re-dimensionalising the conditions \eqref{app:T_inter} yields \eqref{eqn:T_inter}.

The sharp-interface limit will remain valid provided that the radius of solid core of the nanoparticle is much larger than the width of the solid-liquid interface, $\epsilon / R(t) \ll 1$. As the solid core melts, the sharp-interface limit will eventually breakdown, at which point a diffuse-interface model would need to be considered. However, as discussed by Myers \etal\cite{myers2014}, the validity of the continuum limit, which is thought to hold on length scales exceeding 2~nm, also comes into question during nanoparticle melting and might breakdown well before the sharp-interface limit.

\end{appendix}

%-----------------------------------------------------
\section*{References}
\bibliographystyle{elsarticle-num}
\bibliography{refs}

\end{document}